\documentclass[prb, aps,
twocolumn,
 showpacs,amsmath,amssymb]{revtex4}
\usepackage{graphicx}
\usepackage[cp1251]{inputenc}
\usepackage{dcolumn}
\usepackage{bm}
\usepackage{amssymb}
\usepackage{amsmath}
\usepackage{verbatim}
\usepackage{color}
\def \be {\begin{equation}}
\def \ee {\end{equation}}
\def \bea {\begin{align}}
\def \eea {\end{align}}
\def \p {\partial}
\def \BEA {\begin{eqnarray}}
\def \EEA {\end{eqnarray}}

\def \BC {\begin{cases}}
\def \EC {\end{cases}}

\begin{document}

\title{
Rippling and crumpling in disordered free-standing  graphene
}
\author{I. V. Gornyi$^{1,2,3,5}$}
\author{ V. Yu. Kachorovskii$^{1,2,3,5}$}
\author{A. D. Mirlin$^{1,3,4,5}$}
\affiliation{$^{1}$Institut f\"ur Nanotechnologie,  Karlsruhe Institute of Technology,
76021 Karlsruhe, Germany
\\
$^{2}$    A.F.Ioffe Physico-Technical Institute,
194021 St.~Petersburg, Russia
\\$^3$ \mbox{Institut f\"ur Theorie der kondensierten Materie,  Karlsruhe Institute of
Technology, 76128 Karlsruhe, Germany}
\\$^{4}$ Petersburg Nuclear Physics Institute, 188300, St.Petersburg, Russia
\\ $^{5}$
L.D. Landau Institute for Theoretical Physics, Kosygina
  street 2, 119334 Moscow, Russia
}

\date{\today}

\begin{abstract}
 Graphene is a famous realization of elastic crystalline two-dimensional (2D)   membrane.  Thermal fluctuations of  a 2D membrane  tend to destroy the long-range order in the system.  Such fluctuations are  stabilized by strong anharmonicity effects,  which  preserve thermodynamic stability.   The anharmonic effects      demonstrate critical behaviour  on scales larger than the Ginzburg scale.             In particular, clean suspended flake of graphene  shows a power-law increase of the bending rigidity  with the system size,   $\varkappa\propto L^{\eta},$  due to  anharmonic interaction between in-plane and out-of-plane (flexural) phonon modes. We demonstrate  that  random fluctuations of membrane curvature  caused by  static  disorder may change dramatically  the scaling  of the bending rigidity  and lead to a  non-monotonous  dependence of $\varkappa$ on $L.$  We derive coupled renormalization-group equations describing combined  flow of $\varkappa$  and effective disorder strength $b$,
  find a critical curve $b(\varkappa)$ separating  flat  and crumpled   phases,    and  explore  the behavior  of  disorder in the flat phase. Deep in the flat phase,  disorder    decays in a power-law way  at scales larger than the Ginzburg length which therefore   sets a characteristic size for the ripples---static out-of-plane deformations observed experimentally in  suspended graphene.  We  find that in the limit
    $L \to \infty $ ripples are characterized by anomalous exponent $2\eta$ in contrast to dynamical fluctuations governed by $\eta$.
    For sufficiently strong disorder, there exists an intermediate range of spatial
    scales where ripples decay much slower, with exponent $\eta/4$.
    In the near-critical regime, disorder
        first increase  with $L$, then reaches a  maximum  and starts to  decrease. In this case, the
        membrane shows fractal properties implying a multiple folding starting from a certain length scale $L_1$ and finally flattens at a much larger scale $L_2$ (which diverges at criticality).  We conclude the paper by a comparison of our results with available experimental data on graphene ripples.
     \end{abstract}
\maketitle
%%%%%%%%%%%%%%%%%%%%%%%%%%%%%%%%%%%%%%%%%%%%%%%

\section{Introduction}
\label{s1}

Graphene,  a single monolayer of graphite, \cite{Geim,Geim1,Kim} has  attracted enormous
interest in  the last decade (for review, see
Refs.~\onlinecite{geim07,graphene-review,review-DasSarma,review-Kotov,book-Katsnelson,book-Wolf,book-Roche}).
   From the fundamental point of view, this interest  is
largely motivated by quasirelativistic character of its spectrum: charge
carriers in graphene are two-dimensional (2D) massless Dirac fermions. This
leads to a variety of remarkable phenomena.
In particular,  graphene is a unique example of a system where essentially quantum
phenomena such as the quantum Hall effect can be observed up to the room
temperature. \cite{novoselov07}
In view of applications, the technological breakthrough in
fabrication of flat  monolayer   2D crystals
offers new opportunities in the future nanoelectronics.
Remarkably,
suspended graphene  demonstrates  extremely high  room-temperature mobility \cite{Bolotin0, Du,
Bolotin,Du-FQHE,Bolotin-FQHE,Meyer,McEuen,Miao07,Danneau08,tension,Elias,Lau}
as high as  $1.2\cdot10^5$ cm$^2$/Vs   and therefore is considered as  a most
perspective candidate for the carbon-based  nanoelectronics.

Elastic properties of graphene are also quite amazing.     Free-standing  graphene is an outstanding example of an elastic crystalline two-dimensional (2D)   membrane with a high bending rigidity $\varkappa \simeq 1 $ eV.     The most important feature distinguishing such a membrane  from conventional 2D semiconductor systems is the  existence  of
specific type of  the  out-of-plane phonon modes, so called flexural phonons.    \cite{Nelson}
In the harmonic approximation the energy of out-of-plane deformation  reads
\be
    E = \frac{1}{2} \int d{\bf x}\left [{\rho} \dot h^2+ \varkappa (\Delta h)^2\right] ,
\label{energy}
\ee
where  $h({\bf x})$  is  out-of-plane distortion and  $\rho $ is the mass density per unit square.  From Eq.~\eqref{energy} we find  frequency of the flexural phonons
  \be
 \omega_{ \mathbf{q}}= D q^2,
 \label{omega}
 \ee
where $D=\sqrt{\varkappa/\rho}.$
Hence, in contrast to in-plane  acoustic phonons, whose frequency scales as  $q,$  the flexural mode is very soft and, consequently,   the out-of-plane  thermal fluctuations of $h({\bf r})$   are very large.  As a consequence, flexural phonons
 serve as a very effective scattering mechanism for electrons (for discussion of different aspects  of electron-phonon scattering in graphene see Refs.~\onlinecite{Mahan,Suzuura,Hw,Manez,CaKim,Fasolino,aleiner-basko,basko,Oppen-short,Oppen-Scr,
Oppen-long,Vozmediano,Ochoa,kat2,kat1,San-Jose, Ochoa1,my-eph,Hwang13,Blanter,Tikhonov}).

 A remarkable property of flexural phonons is a crucial role played by anharmonic effects.
 In particular,  golden-rule calculation  of scattering  rate  on the deformation  potential created by flexural phonons  leads, in the harmonic approximation and with electrostatic screening taken into account, to values of  the electrical conductivity that are two to three orders of magnitude lower than experimentally observed.  This drastic discrepancy implies  the  existence of a certain mechanism that strongly suppresses  out-of-plane modes.   As was demonstrated in Ref.~\onlinecite{my-eph}, taking into account anharmonic interaction between flexural and in-plane phonons dramatically reduces the electron-phonon scattering, yielding values of  the electrical conductivity that are in good agreement with experimental findings of
  Ref.~\onlinecite {Bolotin}. Hence, a comparison of theoretical results with transport measurements  demonstrates that   anharmonic effects  are extremely  important  in suspended  graphene.

 Suppression of scattering because of  anharmonicity is very favourable for fabrication of ultra-high-mobility graphene structures.
  Further, the anharmonicity governs lattice thermal transport in suspended graphene, which is currently a subject of intense experimental~\cite{Balandin,Seol10,Wang11} and theoretical research (see, e.g. Ref.~\onlinecite{Peeters15} and references therein).
Moreover, anharmonicity plays a key role for the fundamental issue of the  thermodynamic stability of graphene membrane.
Indeed,  due to the softness of flexural-phonon modes they might
be expected to be very efficient in inducing
strong thermal out-of-plane fluctuations and thus  destroying  the membrane \cite{mermin,landau} by  driving  it into the
so-called crumpled phase. \cite{Nelson} This question was intensively discussed in the
literature more than two decades ago
\cite{Nelson,Nelson0,Crump1,NelsonCrumpling,david1,buck,Aronovitz89,david2,lower-cr-D2,d-large,disorders,disorder-imp,Gompper91,
RLD,Doussal,disorders-Morse-Grest,RLD1,Bowick96}
in connection
with biological membranes, polymerized layers and some inorganic surfaces.  The interest to this topic  has been renewed more recently~\cite{eta1,Gazit1,Gazit2,Hasselmann,kats1,kats2,Amorim,kats3}  after discovery of graphene.

It was found \cite{Nelson0,Crump1,NelsonCrumpling,david1,Aronovitz89,buck,david2} that anharmonic coupling of in-plane and out-of-plane phonons
stabilizes the membrane for sufficiently low temperatures $T$, so
that the membrane is in the flat phase at relatively low $T$ and undergoes the
crumpling transition with increasing $T$. The main dimensionless parameter
characterizing the state of the 2D membrane is the ratio of the bending rigidity
$\varkappa$ to the temperature. For graphene this ratio for room temperature is
quite large, $\varkappa/T \simeq 30$. This reflects a remarkable rigidity of
graphene and implies that graphene remains in the flat phase up to the temperatures
several times higher than the room temperature.    Moreover, interaction between flexural  and  in-plane phonons leads to a
power-law renormalization of the bending rigidity\cite{Aronovitz89,lower-cr-D2,Doussal}
\be
\varkappa \to \varkappa(q) \propto {q}^{-\eta},~~\text{for}~~ q \to 0,
\label{bending}
\ee
with a certain critical index $\eta$.
Physically, the increase of the bending rigidity, Eq.~\eqref{bending},  is a
manifestation of the  tendency of the membrane towards the flat phase.

Development of a controllable analytical treatment of renormalization of the bending rigidity and of the crumpling transition is not an easy task.
The central problem is the absence of a small parameter that would control the analysis in the physically relevant case of a membrane  with dimension $D=2$ embedded into the space with dimension $d=3$.  For a membrane with arbitrary $D$ and $d$, a systematic treatment  turns out to be possible in two cases: for $4-D=\epsilon \ll 1$ and for
$d_c=d-D \gg 1$.  In both limits, there exists a small parameter that controls calculations: $\epsilon$ in the  first case, and $1/d_c$     in the second case.  In particular, a theory of crumpling transition for $D=2$ and  $d_c \gg 1$ was developed in Ref.~\onlinecite{david1,david2}, while a renormalization-group (RG) treatment of the membrane elastic coefficients in the limit $\epsilon \ll 1$ was  first discussed in Ref.~\onlinecite{NelsonCrumpling,Aronovitz89}. The value of  the critical exponent  $\eta$ characterising the flat phase was found to be
\be
\eta \simeq  \frac{2}{d_c}
\label{eta00}
\ee
for $D=2$ and $d_c\gg 1$, \cite{david1,david2} and
\be
\eta \simeq \frac{12\epsilon}{24+d_c}
\label{eta000}
\ee
for $\epsilon \ll 1$. \cite{Aronovitz89}

The scaling of the bending rigidity exactly at the crumpling transition point  is characterised by another  critical exponent $\eta_{\rm cr}$.
This exponent determines the fractal (Hausdorff) dimension of the membrane at criticality, $D_H = 2D/(4-D-\eta_{\rm cr})$.
It turns out that for   $D=2$ and $d_c \gg 1$  the exponent $\eta_{\rm cr}$ coincides with $\eta$, Eq.~\eqref{eta00}, in the leading order in $1/d_c$.  On the other hand, in the vicinity of $D=4$  the exponent $\eta_{\rm cr}$  was found to be essentially different from $\eta$, Eq.~\eqref{eta000}, and scaling as $\epsilon^3$, see Refs. \onlinecite{d-large,lower-cr-D2}.

 Further approximate calculation schemes (that become controllable for $d_c\gg 1$)
have been developed, such as the self-consistent screening
approximation (SCSA) \cite{Doussal} and the ``non-perturbative
renormalization group''\cite{eta1}.
After extrapolation to the physical dimensionality, the corresponding results yield $\eta=0.821$ and
$\eta=0.849$, respectively. Clearly, the extrapolation is not controlled
parametrically; the scattering between the above  values may serve as a
rough estimate of their accuracy. Numerical simulations of the problem
gave values  $\eta=0.60 \pm 0.10$
and $\eta=0.72 \pm 0.04$ (see Ref.~\onlinecite{Gompper91} and Ref.~\onlinecite{Bowick96}, respectively).

In the present paper we explore the interplay of a static disorder  and anharmonicities in a graphene membrane. Some aspects of such a problem have been discussed in the literature in a general context of  the membrane physics\cite{disorders,disorder-imp,RLD,disorders-Morse-Grest, RLD1}.
In particular, it was concluded in Refs.~\onlinecite{disorders,disorder-imp,disorders-Morse-Grest} that for $D=4-\epsilon$ with small $\epsilon$
the flat phase is stable with respect to various types of disorder. On the other hand, it was argued in Ref.~\onlinecite{RLD} that, in the leading order in $1/d_c$,  a flat phase of  a  2D  membrane is destroyed already by an infinitely small impurity-induced disorder  (in-plane quenched  random deformations). The later result, if applicable in the physical case $d=3$, would imply an instability of the graphene membrane with respect to an arbitrarily weak disorder. The authors of Ref. ~\onlinecite{RLD} speculated that high-order terms with respect to $1/d_c$ might cure such an instability.  In Ref.~\onlinecite{RLD1}, influence of randomness on the flat phase was treated within the SCSA, with a focus on a long-range disorder.

 Although  effects of disorder in the flat phase (both for $D= 4 -\epsilon$ and $D=2$) were discussed
 in a number of publications, the problem of crumpling transition in a disordered membrane still remains a challenge.
   This problem has a close relation to  formation of ripples which are the  static  out-of-plane random deformations of membrane. In other words,  ripples  look like frozen  flexural phonons.  In graphene,  such deformations with the height $3-10  $ {\AA}  and typical spatial scale about several hundred angstroms   were observed in a number of experiments.\cite{Meyer, Lau,ripp1,ripples1, ripples2,ripples-bao,Kirilenko,ripp2,kir,Neek-Amal,ripp3}   While several recent theoretical papers discussed graphene ripples, \cite{San-Jose,Fasolino,Gazit2,Hasselmann,ripp3} development of a systematic theory of their formation (which would explain the mechanism of rippling and predict key parameters of the emerging structure) remains a challenging problem.

In the present work, we study the  crumpling transition and the rippling in  graphene with a static quenched disorder.
We perform a detailed RG analysis for the out-of-plane (``random curvature'') disorder model.
Previously, the RG treatment of disorder in membranes~\cite{disorders,disorder-imp,disorders-Morse-Grest} has been performed
only for dimensionality $D=4-\epsilon$, which cannot be directly employed to the physical graphene samples.
We assume that the membrane dimensionality $D$ is equal (or close) to 2 and  use the $1/d_c$-expansion which allows us to control the theory and to derive  coupled RG equations describing a combined  flow of $\varkappa$  and effective strength $b$ of the out-of-plane disorder.  In this way, we  establish the phase diagram by determining a critical curve $b(\varkappa)$ separating  the flat and crumpled phases. We demonstrate  that, even deep in the flat phase,  random fluctuation of membrane tension caused by  the static disorder may  strongly change  the critical behavior  of the bending rigidity.   We   discuss in detail  the behavior  of  disorder in the flat phase.

Far from the critical curve, i.e. deep in the flat phase,  disorder    decays in a power-law way  at the scales larger than the Ginzburg length which therefore   sets a characteristic size for the ripples -- static frozen out-of-plane deformations observed experimentally in  suspended graphene. We  find that  in the limit $L \to \infty $  static and dynamic correlation functions of $\nabla h$ behave as $L^{-2\eta}$ and $L^{-\eta},$ respectively.
Hence, ripples are characterized by the anomalous exponent $2\eta$ in contrast to dynamical fluctuations governed by $\eta.$
Furthermore, we show that, for sufficiently strong disorder, there exists an intermediate range of spatial
    scales where ripples decay much slower, with exponent $\eta/4$.
   One of remarkable results is that the RG flow of coupling constants may be essentially non-monotonous. Specifically, for sufficiently strong disorder (close to the critical curve), the bending rigidity decreases at the first stage of the renormalization,   reaches its minimum,  and  then starts to grow.  In this  near-critical regime  the  disorder also changes in a non-monotonous fashion: it  first increases very slowly (logarithmically) with the spatial scale $L$, then reaches a  maximum at  a certain scale $L_{ r} $ and, finally, decreases  according to  a power law at larger scales. In this case, the membrane shows fractal properties which imply its multiple folding starting form a certain length scale $L_1$ and flattens  at a much larger scale $L_2$.

 We also  briefly discuss an in-plane disorder and show that it is irrelevant in the RG sense, unless its correlation function is highly singular at small momenta, i.e., unless it is too long-ranged. Thus, the above conclusions remain valid also in the presence of in-plane disorder. The in-plane disorder may, however, affect essentially the bare value of the out-of-plane disorder at the atomic scales, which serves as a starting point for the RG analysis.

We conclude the paper by a comparison of our results with available experimental data on graphene ripples. While the main focus of the paper is on graphene, the developed theory is quite general and is expected to be applicable to other crystalline membranes as well.  These include biological membranes like those of red blood cells,~\cite{biomembrane} oxidised graphite or graphene,~\cite{graphite-oxide} graphane,~\cite{Costamagna} molybdenum disulphide,~\cite{MoS2} and boron nitride \cite{BN} membranes. Further examples may likely emerge in near future, in view of current active works on engineering of 2D materials and structures.

\section{Formulation of the problem and  mean-field analysis}

To begin with, we note that the  energy of membrane consists of kinetic and elastic contributions.   In this paper, we treat the problem semiclassically (see Appendix~\ref{quasiclassical} for discussion of the region of applicability of the quasiclassical approximation).    The     kinetic energy depends  on  momenta only.    Within the semiclassical approximation, the phonon momenta can thus be integrated out from the very beginning.  In what follows we neglect the kinetic term (see Refs.~\onlinecite{kats1,kats2,kats3} for a discussion of some effects related to this term) and focus on the study of the elastic terms.

\begin{figure}[t]
%\center
\centerline{\includegraphics[width=0.4\textwidth]{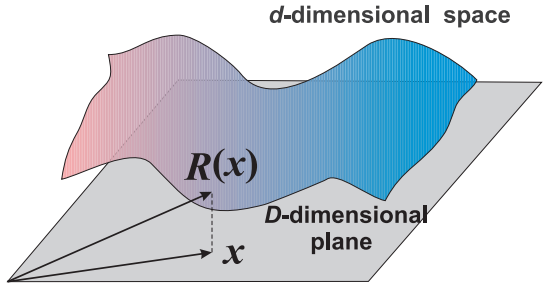} }
\caption{
  Membrane  with dimension $D$ is embedded into a  space with dimension $d.$    A  point on the membrane   surface is  labeled by a   $d-$dimensional vector $\mathbf R$   which depends on
vector $\mathbf x$ belonging to an arbitrary $D-$dimensional  reference plane.}
\label{Fig1}
\end{figure}

 We start from  the clean case and consider  a general Landau-Ginzburg form  of the  elastic free energy: \cite{NelsonCrumpling}
\BEA
&& F = \int d^Dx \left\{ \frac{w }{2} (\p_\alpha \p_ \alpha \mathbf R )^2
-\frac{t}{2} (\p_\alpha \mathbf R \p_\alpha \mathbf R ) \right. \label{Fstart} \\ \nonumber
 && \left. + u
(\p_\alpha \mathbf R \p_\beta \mathbf R )^2 +v (\p_\alpha \mathbf R \p_\alpha \mathbf R )^2
 \right \} ,
\EEA
Here $D$ is the dimension of the membrane, $d$ is the dimension of the embedding space and $\alpha,\beta=1,...,D$.  In principle,  Eq.~\eqref{Fstart} allows one to describe both flat and crumpled phases of the membrane.  In the latter case, one should include an additional term that prevents self-intersections.  In this paper, we focus on the flat phase, so that we omit this term.
The  $d-$dimensional  vector $\mathbf R$ depends on the point of the $D$-dimensional reference space, $\mathbf x=(x_1,...,x_D)$, i.e.,
$\mathbf R=\mathbf R (\mathbf x)$ (see Fig.~\ref{Fig1}).  Equation (\ref{Fstart}) can be obtained by following considerations. First, the translational invariance implies that the free energy depends on $\mathbf R$ via the derivatives $\p_\alpha \mathbf R$ only. Second, the rotational invariance requires that the energy should be scalar with respect to both the embedding ($\mathbf R$) space and the reference ($\mathbf x$) space. Finally, keeping the leading terms in the expansion in the field $\p_\alpha \mathbf R$ and in gradients yields Eq.~(\ref{Fstart}).

Following Ref.~\onlinecite{NelsonCrumpling},  we first  consider the mean field approximation which amounts to the linear ansatz,
$$\mathbf R=\xi_0 \mathbf x.$$
This yields the free energy of Landau type,
\be
F\propto -\xi_0^2 t+2\xi_0^4 (u+Dv).
\label{landau-free-energy}
\ee
Next, we find $\xi_0$ by minimization of the   free energy (\ref{landau-free-energy}):
\be  \p F/\p\xi_0=0 ~\Rightarrow ~\xi_0^2= \left\{\begin{array}{c}
                                             \displaystyle\frac{t}{4(u+Dv)}, ~\text{for }~t>0 \\\\
                                             \hspace{5mm}0, \hspace{7mm}\text{for }~t<0
                                           \end{array}
 \right.
 \label{mean}
 \ee
 The case with  $\xi_0^2>0$ corresponds to the flat phase, while for  $\xi_0^2=0$  the manifold  $\mathbf R (\mathbf x)$  shrinks to a point  $\mathbf R=0$, which  implies
 that the membrane is crumpled. (Taking self-avoidance into account would lead to a finite radius of the crumpled membrane.\cite{NelsonCrumpling})
Assuming (in spirit of Landau theory of phase transitions) that $t\propto T_c-T,$ we find that    $\xi_0^2 \propto T_c-T.$  Hence, the  model Eq.~\eqref{Fstart}  shows a    crumpling  transition at $T=T_c$ already within the mean-field approximation.

%%%%%%%%%%%%%%%%%%%%%%%%%%%%%%%
%%%%%%%%%%%%%%%%%%%%%%%%%%%%%%%%%

We  assume now that the system is in the flat phase  $T< T_c$ and take into account thermal fluctuations around the mean-field solution.
For this purpose, we write
\be \mathbf R=\xi_0 \mathbf r,
\label{Rmean}
\ee
where
\be
\mathbf r=\mathbf x + \mathbf u+\mathbf h
\label{xsi=1}
\ee
and vectors
$$\mathbf u=(u_1,...,u_D),~ \mathbf h =(h_1,...,h_{d-D}) $$
represents in-plane and out-of-plane displacements, respectively.
Substituting  Eq.~\eqref{Rmean} into Eq.~\eqref{Fstart} and choosing for $\xi_0$ the mean-field solution Eq.~\eqref{mean}, we find
\BEA
&&F=\int d^Dx \left\{ \frac{\varkappa }{2} (\Delta \mathbf r )^2     \right.  \label{Fr}\\ \nonumber
&& \left.+\frac{\mu }{4} [\partial_\alpha \mathbf r \partial_\beta \mathbf r-\delta_{\alpha\beta} ]^2+\frac{\lambda }{8}
[\partial_\gamma \mathbf r \partial_\gamma \mathbf r-D ]^2
\right\}.
\EEA
Here
\be
\varkappa = w \xi_0^2
\label{kappa-xi^2}
\ee
is the bending rigidity  and $$ \mu=4u\xi_0^4,~\lambda=8v\xi_0^4 $$  are
in-plane elastic constants.
We see that these constants as well as  $\varkappa$ turn to zero at the transition point:
$\mu,\lambda \propto (T_c-T)^2,~\varkappa \propto T_c-T .$

Next, we use the parametrisation \eqref{xsi=1} and rewrite  Eq.~\eqref{Fr} in terms of $\mathbf u$ and $\mathbf h.$
   We assume that   spatial derivatives of fields are small,       $|\p_\alpha \mathbf h |\ll 1,$   $|\p_\alpha \mathbf u| \ll 1,$  so that one can neglect terms  $  \p_\alpha \mathbf u  \p_\beta \mathbf u$ compared to $\p_\alpha \mathbf u$ in the second and third contributions to the energy \eqref{Fr}.    Further, one may   neglect  the  term $\varkappa (\Delta \mathbf u)^2 $ compared to $\mu (\p_\alpha \mathbf u_\beta)^2$  and $  \lambda (\p_\alpha \mathbf u_\alpha)^2$ provided that the  characteristic  length scale  of the  variation of the membrane displacement, $q^{-1},$  is  sufficiently large:
\be
q \ll  \rm{min} \left(\sqrt{\frac{\mu}{\varkappa}}, \sqrt{\frac{\lambda}{\varkappa}}\right).
\label{grad}
\ee
Under these assumption,
 one can rewrite  Eq.~\eqref{Fr} for membrane energy in the standard  textbook form \cite{Nelson}:
 \be
 F = \int d^Dx\left\{  \frac{\varkappa}{2} (\Delta\mathbf h)^2+\mu u_{ij}^2+\frac{\lambda}{2}u_{ii}^2   \right\},
 \label{Fstandard}
 \ee
where
\be
u_{\alpha\beta}=\frac{1}{2}\left(\p_\alpha \mathbf r\p_\beta \mathbf r-\delta_{\alpha\beta} \right) \approx \frac{1}{2}\left(\p_\alpha u_\beta
+
 \p_\beta u_\alpha +\p_{\alpha} \mathbf h \p_\beta\mathbf h  \right)
 \label{uab}
 \ee
is the deformation tensor.

 The following comment is in order here.  The question about the range of  applicability of the simplified model, Eq.~\eqref{Fstandard} is in fact somewhat more subtle than one might conclude from the above discussion. The point is that the  elastic constants $ \mu$, $\lambda$, and $\varkappa$   are strongly renormalized   due to anharmonicity and critical fluctuations that it induces.
This effect becoming prominent for  $q\ll q^*$, where $q^*$ is the inverse Ginzburg scale, see Eq.~ \eqref{q*} below. At such scales, the criterion \eqref{grad}  for neglecting the $\varkappa (\Delta \mathbf u)^2 $ term should be modified.  In particular, for $D=2,$  the screening by critical fluctuations reduces $\mu$ and $\lambda$ to the value   $ \sim T/\Pi_\mathbf q\sim \varkappa ^2 q^2/T$ (see section \ref{clean}).  Hence, $q$ drops out from the estimate Eq.~\eqref{grad} and instead we get the condition $\varkappa \gg T. $ As will be demonstrated below, the crumpling transition corresponds to $\varkappa \sim d_c^2 T, $ where $d_c=d-D$.  Hence, the simplified  model \eqref{Fstandard} is sufficient for a description of the crumpling transition  provided that $d_c \gg 1$. On the other hand, it is exactly the latter condition that we will use to develop a controllable RG treatment for the theory \eqref{Fstandard}. Therefore, neglecting the $\varkappa (\Delta \mathbf u)^2 $ term and thus restricting ourselves with the free energy \eqref{Fstandard} is fully consistent with the subsequent RG analysis.

Let us now  introduce  a quenched disorder into the model \eqref{Fstandard}.
We will mainly focus on the out-of-plane (random curvature) disorder in this paper.~\cite{disorders,disorders-Morse-Grest}
Our motivations for analysing the effect of this type of disorder are twofold.
First, we will show that the random-curvature disorder influences the RG flow (and thus the crumpling transition) in an essential way. This should be contrasted with the in-plane disorder, which is RG-irrelevant (unless its correlation function is highly singular at low momenta), as will be shown below. Second, such disorder induces static out-of-plane deformations and is thus directly related to the effect of rippling.

The free energy including the random-curvature disorder reads~\cite{disorders}
\be
F (\mathbf u,\mathbf h)= \int d^Dx\left\{  \frac{\varkappa}{2} (\Delta\mathbf h +\boldsymbol{\beta})^2+\mu u_{ij}^2+\frac{\lambda}{2}u_{ii}^2   \right\}
\label{Fdis} .
\ee
Here $\boldsymbol{\beta}=\boldsymbol{\beta(\mathbf x)}$ is a random vector with  Gaussian distribution
\be
P(\boldsymbol{\beta}) =Z_{\boldsymbol{\beta}}^{-1}\exp\left( -\frac{1}{2b} \int \beta^2(\mathbf x) d^D\mathbf x \right),
\label{dis}
\ee
where $b$ is the disorder strength and  $ Z_{\boldsymbol{\beta}}$ is a normalization factor.
In order to study  the membrane properties in a systematic way, one should take  into account fluctuations,
thus going beyond the mean-field approximation.

 \section{Beyond mean field}
 \label{sec:beyond-mf}

In this Section, we develop an RG treatment of a disordered membrane.
Previously, the curvature disorder was treated within the RG approach~\cite{disorders-Morse-Grest} only in dimensionality $D=4-\epsilon$.
We will see below that  the structure of the RG near $D=2$ is crucially different from the case $D=4-\epsilon$.
A lowest-order perturbative treatment of the curvature disorder in physical dimension $D=2$ was briefly discussed in
Ref.~\onlinecite{disorders}. However, this consideration is only applicable for short distances (smaller than then the Ginzburg length), whereas the most interesting physics (captured by the RG) developes on larger scales.

 Our analysis is based on an expansion around the ordered (flat-membrane) phase, and thus bears  analogy to a $\sigma$-model treatment of conventional critical phenomena. In full similarity with the $\sigma$-models, $D=2$ is a logarithmic dimension for the present problem (which manifests in  $\varkappa/T$ becoming dimensionless in this case), so that our analysis will be appropriate at or near $D=2$. This is highly favourable, since  $D=2$ is exactly the physical dimension of a membrane. To keep the theory under control, we will also assume a large dimensionality of the embedding space, $d_c \gg 1$.

 \subsection{Renormalization of stretching parameter $\xi$}

 Above we rescaled coordinates with the scaling factor  $\xi_0$ which  minimize the free energy within the mean-field approximation.   Beyond mean field   one should  take into account critical fluctuations. Such fluctuations change the optimal scaling factor.  To consider  this effect, we modify Eq.~\eqref{xsi=1}  as follows:
\be \mathbf r=\xi \mathbf x+ \mathbf u +\mathbf h.  \label{xi-new1}\ee
 Equation \eqref{xi-new1} represents a new rescaling of coordinates.  The mean-field approximation corresponds to $\xi=1.$   Below,   we  find  that $\xi$ flows  away from the mean-field value  due to the fluctuations.
 Substituting Eq.~\eqref{xi-new1} into Eq.~\eqref{Fr} and using the same approximations that were done in the course of derivation of Eq.~\eqref{uab},  we arrive at the following expression for the free energy:
 \BEA
 F&=&\frac{D L^D (\mu+\lambda D/2)}{4} \left[ (\xi^2-1)^2 \right. \label{zero mode}
  \\  &+& \left. \frac{2(\xi^2-1)}{D}~ \int  \frac{ d^D\mathbf x}{L^D}~  \p_\alpha \mathbf h  \p_\alpha \mathbf h
  \right]
+F(\tilde {\mathbf u}, \mathbf h),
\nonumber
 \EEA
 where
$ F(\tilde {\mathbf u}, \mathbf h)$ is given by  Eq.~\eqref{Fdis}  with $\mathbf u$  replaced by $\tilde{\mathbf u} = \xi\mathbf u$.

 Since the   product $  {\p_\alpha \mathbf h  \p_\alpha \mathbf h } $  in Eq.~\eqref{zero mode}
  is averaged over the  whole volume  of the  membrane, its fluctuations are negligibly  small  in the limit $L \to \infty$.       Hence, one can replace
 $$  {\p_\alpha \mathbf h  \p_\alpha \mathbf h  } \to  \left \langle   {\p_\alpha \mathbf h  \p_\alpha \mathbf h}  \right \rangle, $$
  where $\langle \cdots \rangle$ stands for averaging with the statistical factor $\exp[-F(\tilde {\mathbf u}, \mathbf h)/T].$   Minimizing thus obtained energy with respect to $\xi$,  we find that the optimal value of $\xi$ deviates from the mean field value $\xi=1$ due to the fluctuations,
  \be
  \xi^2=1- \frac{1}{D}
    ~ \left \langle \p_\alpha \mathbf h  \p_\alpha \mathbf h \right \rangle.
  \label{dxi}
  \ee
In order to calculate  $\left \langle   {\p_\alpha \mathbf h  \p_\alpha \mathbf h}  \right \rangle,$ we notice that  $F(\tilde {\mathbf u}, \mathbf h)$ contains linear, quadratic, cubic, and quartic   terms with respect to variables $\mathbf h $ and  $\mathbf u.$   In this section, we neglect  cubic and  quartic terms, thus neglecting anharmonicity.  The anharmonicity-related  effects will be included in the next section in the framework of the RG formalism.
In the harmonic approximation, the $\mathbf u$-dependent terms in the free energy do not couple with   $\mathbf h$-dependent ones and, therefore, can be integrated out from the very beginning.    The  $\mathbf h$-dependent part of the energy  in the harmonic  approximation is given by
$\int d^Dx \: \varkappa (\Delta\mathbf h +\boldsymbol{\beta})^2/2.$   An averaging over fluctuation of $\mathbf h$ and over disorder  yields
\be
\left \langle   {\p_\alpha \mathbf h  \p_\alpha \mathbf h}  \right \rangle = d_c \left( \frac{T}{\varkappa} +b\right) \int \frac{d^D \mathbf q}{ (2\pi)^D } \frac{1}{q^2}.
\label{hhh}
\ee
For $D =2$
 the integral diverges logarithmically. We will see below that $D =2$ is the lower critical dimension of the theory in the limit of infinite dimensionality of the embedding space $d\to\infty$. The special role of $D =2$ is not unexpected, since the theory that we are developing is based on an expansion near the ordered phase ($\xi >0$) and thus bears analogy with $\sigma$-models (cf.   Ref.~\onlinecite{david1}).  Also, this role of $D =2$
 can be foreseen already after a brief inspection of the free energy (\ref{Fdis}),  (\ref{dis}), since both coupling constants $T/\kappa$ and $b$ have a dimensionality of  $L^{D-2}$ and thus become dimensionless in 2D.

 In a vicinity of  the logarithmic dimension $D =2$, an RG formalism can be developed in the conventional way.
    Introducing an  infrared cutoff $k$,  making change of variables
  \be
  \tilde{\xi}^2= \xi^2  k^{2-D},
  \ee
     and using Eqs.~\eqref{dxi}  and \eqref{hhh}, we find  an  RG equation that determines a flow of the stretching parameter $\xi$ with the  spatial scale $k^{-1}$,
\be
\frac{d \tilde{\xi}^2}{ d\Lambda} =  (D-2)\tilde {\xi}^2 - \frac{\tilde {d}_c}{4\pi}\left(\frac{T}{\varkappa} + b \right),
\label{RGxixi}
\ee
where $\Lambda=\ln\left( k_m/ {k} \right) $ and
\be
\tilde{d}_c= \frac{d_c}{ (4\pi)^{(D-2)/2}\Gamma(D/2+1)}.
\ee
Exactly at    $D=2,$   Eq.~\eqref{RGxixi} simplifies to
\be
\frac{d {\xi}^2}{ d\Lambda} =   - \frac{{d}_c}{4\pi}\left(\frac{T}{\varkappa} + b \right),  \qquad D = 2.
\label{RGxixi2D}
\ee
The renormalization of $\xi$ is terminated by $k=1/L$.
The first and second  terms in the r.h.s.  of Eq.~\eqref{RGxixi2D}  describe contributions of dynamic fluctuations (flexural phonons) and of static deformations (ripples), respectively.
Equation  \eqref{RGxixi2D} predicts that   $\xi$ becomes zero at a finite system size. In other words,  a 2D  membrane should be  crumpled in the thermodynamic limit due to both dynamic  and static deformations. While this conclusion is reminiscent of the Mermin-Wagner theorem\cite{mermin} that forbids a long-range order in 2D systems, it turns out to be wrong (which is a manifestation of the fact that the Mermin-Wagner theorem is not applicable to the problem under consideration\cite{david1}). Specifically, inclusion of anharmonicity leads to renormalization of $\varkappa$ and $b$ and restores the ordered (flat) phase in 2D for a finite dimensionality of the embedding space, $d<\infty$.

We are now going to derive  RG equations describing the renormalization of elastic constants and disorder. We will see that  at sufficiently large spatial scales, $\mu$ and $\lambda$ are  screened   to values   independent  on the bare  ones. The effective theory at such scales is thus determined by three running couplings: the stretching parameter (``field renormalization'') $\xi$,  the  bending rigidity (``stiffness'') $\varkappa$  and  the effective  disorder strength $b$.   We start from a discussion of  renormalization of $\varkappa$  in the clean case ($b=0$) in Sec.~\ref{clean} and then derive the full set of RG equation  for disordered membrane in Sec.~\ref{disordered} .

\subsection{Renormalization of the bending rigidity in the clean case }

\label{clean}

We start from the clean-membrane free energy $F(\tilde{\mathbf u }, \mathbf h) $. Since the theory is quadratic with respect to longitudinal modes $\mathbf u$, they can be integrated out.\cite{Nelson0,Doussal}    The stretching parameter $\xi$ enters $F(\tilde{\mathbf u }, \mathbf h) $  via $\tilde{\mathbf u}=\xi \mathbf u$ only, and evidently drops out after  changing variables $\mathbf u \to \xi \mathbf u$  in the    functional integral over $\mathbf u.$       Hence, we  arrive at the free energy   that depends on $\mathbf h$ vectors only: \cite{Doussal}
\BEA
\frac{F}{T}&=& \frac{\varkappa}{2T}\int (dk) k^4 |\mathbf h_\mathbf k|^2 \label{F} \\ \nonumber
&+&\!\! \frac{1}{4d_c}\!\int (dk dk' dq) R_\mathbf q(\mathbf k, \mathbf k') \left( \mathbf h_{\mathbf k+\mathbf q} \mathbf  h_{-\mathbf k} \right) \left( \mathbf h_{-\mathbf k'-\mathbf q} \mathbf  h_{\mathbf k'} \right).
\EEA
Here and below we use a short-hand  notation $(dk)=d^D\mathbf k/(2\pi)^D.$
The kernel of  quartic interaction between transverse modes (see Fig. \ref{Fig2})  can be cast in the form (see Appendix \ref{technical})
\be
R_\mathbf q(\mathbf k, \mathbf k') ={\cal N} \frac{\mathbf k_\perp^2\mathbf k_\perp^{'2}}{D-1} +{\cal M} \left[
\left(\mathbf k_\perp\mathbf k_\perp^{'}\right)^2 -\frac{\mathbf k_\perp^2\mathbf k_\perp^{'2}}{D-1}\right].
\label{K}
\ee
Here $$\mathbf k_\perp= \hat P \mathbf k=\mathbf k- \frac{\mathbf q (\mathbf k\mathbf q)}{q^2},$$
 where $\hat P$ is the  projection  operator related to the transferred momentum $\mathbf q$,
\be
P_{\alpha\beta} =\delta_{\alpha\beta}- q_\alpha q_\beta/q^2,
\ee
and
\be
{\cal N}=\frac{\mu(2\mu+D\lambda)}{(2\mu +\lambda)T}, \qquad {\cal M}=\frac{\mu}{T}.
\ee
Hence, for a generic dimensionality $D$ of the membrane, there are two coupling constants, $\cal N$ and $\cal M$, controlling the interaction strength.   An important exception is the case $D=2$ where the constant ${\cal M}$ drops out from the theory because the term in the square brackets in Eq.~\eqref{K} turns to zero. Hence, for $D=2$ the interaction reads
\be
R^{D=2}_\mathbf q (\mathbf k,\mathbf k')=\frac{2\mu (\mu+\lambda)}{(2\mu +\lambda)} \frac{[\mathbf k\times \mathbf q]^2}{q^2}\frac{[\mathbf k'\times \mathbf q]^2}{q^2}.
\label{R2D}
\ee
The bare  propagator (which is exact in the absence of interaction, ${\cal N}={\cal M}=0$) is given by
\be
\langle  h_{ \mathbf k}^{ \alpha} h_{ - \mathbf k'}^{ \beta} \rangle=  (2\pi)^D\delta(\mathbf k-\mathbf k')~\delta_{\alpha\beta}~G_\mathbf k^0,
\label{hh}
\ee
where
\be
G_\mathbf k^0  = \frac{T}{\varkappa k^4}.
\label{G00}
\ee

The interaction coupling constants get screened  in analogy with conventional charges  in a media with a finite polarizability.
Evaluating the  polarization operator to the  one-loop order [which is nothing but the random phase approximation (RPA)],
we find the screened couplings:
    \begin{eqnarray}
      {{\cal N}_\mathbf q}  &=& \frac{{\cal N}}{1+{\cal N}(D+1)\Pi_\mathbf q},
      \label{N}\\
       {{\cal M}_\mathbf q}  &=& \frac{{\cal M}}{1+2{\cal M}\Pi_\mathbf q}.
       \label{M}
    \end{eqnarray}
Thus, the coupling constants $\cal N$ and $\cal M$ become $\mathbf q$-dependent  and  are screened independently of each other. There is, however, an invariant subspace of the elastic coefficients,
\be \lambda=-\frac{2\mu}{D+2},  \label{invariant} \ee
  where the coupling constants  stay equal up to a numerical coefficient:
  \be {\cal N}_\mathbf q= \frac{2{\cal M}_\mathbf q} {D+1}. \label{NMinvar}
  \ee
The polarization operator $\Pi_\mathbf q$ reads
\be
\Pi_\mathbf q=\frac{1}{D^2-1} \int (dk) k_\perp^4 G_\mathbf k^0 G_{\mathbf q-\mathbf k}^0.
\label{Pi}
\ee
Using (\ref{G00}), we get
 \be
 \Pi_\mathbf q = A_D \frac{T^2 }{\varkappa ^2 q^{4-D}}
 \label{Pi0}
 \ee
 where
\be
A_D=\frac{\Gamma\left( \frac{D}{2}\right) \Gamma\left( \frac{4-D}{2}\right)  }{\pi ^{(D-1)/2} ~2^{2D+1}~ \Gamma\left( \frac{D+1}{2}\right) }.
\ee
For $D=2,$ we get $A_2=1/16\pi.$

%%%%%%%%%%%%%%%%%%%
\begin{figure}[t]
\centerline{\includegraphics[width=0.45\textwidth]{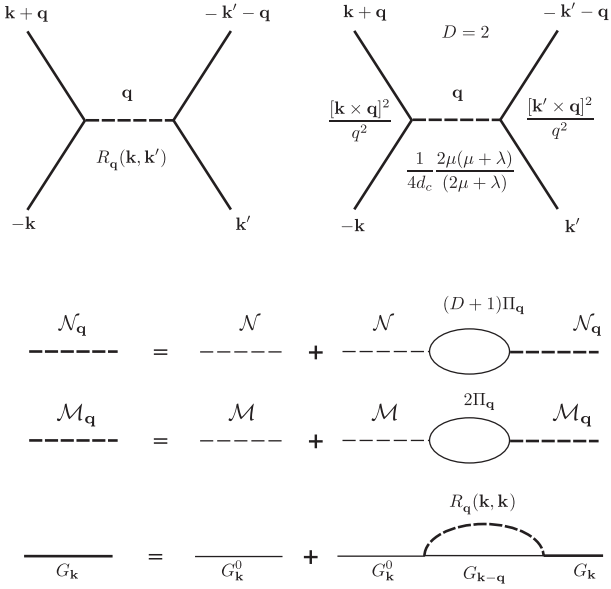} }
\caption{{\it Top:} $h^4-$ interaction vertex for generic $D$ (left) and for $D=2$ (right).   {\it Bottom:}  One-loop diagrams
describing renormalization of the interaction and of the propagator in a clean membrane.
}
\label{Fig2}
\end{figure}
%%%%%%%%%%%%%%%%%%%%%

It follows from Eq. ~\eqref{Pi0} that for any $D<4$ the polarization operator increases according to a power law with decreasing $q$.  Hence, as seen from Eqs.~\eqref{N} and \eqref{M}, for  sufficiently small $q$  couplings $\cal N$ and $\cal M$  become independent of their bare values and inversely proportional to   the polarization operator:
\be
 {\cal N}_\mathbf q \approx\frac{1}{(D+1)\Pi_\mathbf q},   \qquad {\cal M}_\mathbf q\approx \frac{1}{2\Pi_\mathbf q}.
 \label{NMscreened}
 \ee
For the invariant subspace \eqref{invariant},     Eqs.~\eqref{NMscreened} hold  for $q \ll q^*,$ where
\be
q^* \sim\left(\frac{\mu T}{ \varkappa^2}\right)^{1/(4-D)}
\label{q*}
\ee
is the inverse Ginzburg length  which separates the normal region ($q>q^*$) from the critical one  ($q<q^*$).
For the case when bare couplings ${\cal N}$ and ${\cal  M}$ are essentially different, there are two different scales $q^*_{\cal N}$ and  $q^*_{\cal M}$  at which   they become screened according to Eq.~\eqref{NMscreened}.
For simplicity, we will assume that the bare values are of the same order, ${\cal N} \sim {\cal M}$, so that  $q^*_{\cal N} \sim q^*_{\cal M} \sim q^*$. For a 2D membrane this question does not even arise, since the coupling ${\cal M}$ is simply absent.

We are now ready to evaluate the renormalization of the bending rigidity by (screened) interaction. The renormalized propagator of the $h$-field is given by
\be
 G_\mathbf k = \frac{T}{\varkappa k^4 +\Sigma_\mathbf k } \label{G},
 \ee
where the one-loop self-energy reads
\be
\Sigma_\mathbf k=\frac{2T}{d_c}\int (dq)\frac{{{\cal N}_\mathbf q}+(D-2){{\cal M}_\mathbf q}}{D-1}k_\perp^4G^0(\mathbf k-\mathbf q).
\label{Sigma}
\ee
For $q \ll q^*,$  the effective
%constant
screened interaction
$[{\cal N}_\mathbf q+(D-2){\cal M}_\mathbf q]/(D-1)$  that enters   Eq.~\eqref{Sigma} becomes
\be
\frac{{{\cal N}_\mathbf q}+(D-2){{\cal M}_\mathbf q}}{D-1}\approx \frac{D}{2 (D+1)\Pi_\mathbf q}.
\label{NM}\ee
Let us now substitute Eq.~\eqref{NM} into Eq.~\eqref{Sigma}  and consider the behavior of $\Sigma_\mathbf k$ at low momenta $k \ll q^*.$
Using Eq.~\eqref{Pi0} for the polarisation operator, we find that the integral in Eq.~\eqref{Sigma} scales  as $k^4\ln(1/k)$. This implies a logarithmic renormalization  of the bending rigidity:
\begin{eqnarray}
\delta \varkappa (\mathbf k)  &\simeq& \frac{D\varkappa}{d_c(D+1) A_D} \int\limits_0^{q^*} (dq) \frac{k_\perp^4 q^{4-D}}{k^4|\mathbf k-\mathbf q|^4}\nonumber\\
&\simeq &  \eta \varkappa \ln\left(  \frac{q^*}{k}\right) ,
\label{dkappa}
\end{eqnarray}
 where we neglected the   contribution from  $q>q^*$ which  is convergent for $k\to 0$ and is therefore  small.  The constant $\eta$ can be expressed as
 \be
  \eta= \frac{1}{d_c}\frac{D}{D+1} \left[ \frac{B(D,\eta_0) \eta_0}{A(D,\eta_0)}\right]_{\eta_0\to 0},
\label{eta0}
\ee
with integrals  $A(D,\eta)$ and  $B(D,\eta)$ as defined in  Appendix \ref{regul}. The functions  $A(D,\eta)$ and $B(D,\eta)$  will be used below for comparison with  SCSA.\cite{Doussal}  Note that $A(D,0)=A_D.$

For a sufficiently large spatial scale $k^{-1}$ the correction \eqref{dkappa} ceases to be small and gets promoted to a RG equation,
\be
 \frac{d\varkappa}{d \Lambda}=  \eta  \varkappa,
\label{kappa0}
 \ee
 where $\Lambda= \ln(q^*/k)$. We thus see that $\eta$ is the anomalous dimension of the bending rigidity.
Using Eqs.~\eqref{A} and \eqref{B}, we find
\be
 \eta= \frac{D(D-1) 2^D}{\sqrt \pi d_c}
 \frac{\Gamma\left( \frac{D+1}{2}\right)}{ \Gamma\left(\frac{ D}{2}\right) \Gamma\left(2- \frac{D}{2}\right)\Gamma\left( 2+\frac{D}{2}\right)}.
\label{eta}
\ee
It is worth pointing out that, at variance with Eq.\eqref{hhh},  the integral  \eqref{dkappa} is always logarithmic (i.e., also away from $D=2$). Thus, Eq. \eqref{eta} is applicable for any $D$ as well. The only
condition of validity of this equation is $d_c\gg 1$, which implies that  $\eta \ll 1$.   The equation simplifies for $D=2$  and  $D=4-\epsilon$, yielding
\be
\eta  \simeq \left\{  \begin{array}{c}
         {2}/{d_c},  \hspace{3mm}\text{for}~D=2, \vspace{2mm}\\
          {12\epsilon}/{d_c},  \hspace{3mm}\text{for}~D=4-\epsilon.
       \end{array} \right.
\label{eta2}
\ee
[Since $d_c\gg 1,$  the second line in this equation agrees with Eq.~\eqref{eta000} obtained  by $\epsilon-$expansion.]
 For generic values of  $d$ and $D$,   the exponent $\eta$ is not small. To find it, one would have to take into account higher-loop contributions to RG equations.  Such a calculation does not appear feasible because of the absence of a small parameter.  On the other hand, one may develop a self-consistent extension of the one-loop theory by inserting renormalized Green functions  into the one-loop diagrams, which amounts to the replacement~\cite{Doussal} of $G^0$ with $G$ in Eqs.~\eqref{Pi} and \eqref{Sigma}.   Corresponding calculations yield a self-consistency equation for $\eta$
 \be
   1= \frac{1}{d_c}\frac{D}{D+1}  \frac{B(D,\eta) }{A(D,\eta)},
\ee
 which was derived in Ref.~\onlinecite{Doussal}.  This equation can be  obtained  from Eq.~\eqref{eta0} by replacing $\eta_0$ with $\eta$.
 Although such a procedure is not controlled parametrically for physical membranes ($d=3$, $D=2$),
 it gives  in this case  a value $\eta \approx 0.821$ which turns out to be in a reasonable  agreement~\cite{foot-Gazit}
 with numerical values  $\eta=0.60 \pm 0.10$
and $\eta=0.72 \pm 0.04$ (see Ref.~\onlinecite{Gompper91} and Ref.~\onlinecite{Bowick96}, respectively).

\subsection{Renormalization group for  disordered membrane }
\label{disordered}

Now we include  in the consideration the random curvature disorder which modifies  only the first term in Eq.~\eqref{F}.         In the coordinate representation,  this term becomes
\be
F_0=\frac{\varkappa}{2}\int d^D\mathbf x \left( \Delta \mathbf  h +\boldsymbol{\beta} \right)^2,
\label{F0}
\ee
where $\boldsymbol{\beta}=\boldsymbol{\beta(\mathbf x)}$ is a random vector with  Gaussian distribution (\ref{dis}).
Disorder averaging can be performed with making use of the replica trick.  To this end, we  introduce $N$ replicas of  the field $\mathbf h$  (i.e., make a replacement $\mathbf h \to \mathbf h^n$ with a replica index   $n=1, \ldots, N $) and  replicate  the free energy $F$:
   \be
   F^{\rm{rep}}= F^{\rm{rep}}_0+ F^{\rm{rep}}_1,
   \ee
   where
   \be
   F^{\rm{rep}}_0=
 \sum\limits_{n=1}^{n=N}  \frac{\varkappa}{2}\int (dk) k^4 |\mathbf h_\mathbf k^n +\boldsymbol{\beta}_\mathbf k  |^2
\ee
and
\BEA
 &&F^{\rm{rep}}_1 = \sum\limits_{n=1}^{n=N} \frac{1}{4d_c} \label{F-repl}\\
 &&\times \int (dk dk' dq) R_\mathbf q(\mathbf k, \mathbf k') \left( \mathbf h_{\mathbf k+\mathbf q}^n \mathbf  h_{-\mathbf k}^n \right) \left( \mathbf h_{-\mathbf k'-\mathbf q}^n \mathbf  h_{\mathbf k'}^n \right).
\nonumber
\EEA
 Next, we average $\exp(-F_{\rm{rep}}/T)$ with $P(\boldsymbol{\beta} )$,   thus arriving at the following effective action:
\be
F_{\rm{eff}}= \frac{1}{2} \sum\limits_{n,m} \int (dk) \varkappa^{nm} k^4 \mathbf h_\mathbf k^n \mathbf h_{-\mathbf k}^m + F^{\rm{rep}}_1,
\ee
where  we have introduced a replica-space matrix $\hat {\varkappa}$ with elements   $\varkappa^{nm}$ given by
\be
\hat{ \varkappa}=\varkappa   - \frac{b\varkappa^2 }{T+b\varkappa N} \hat J.
\label{kappa-matrix}
\ee
Here  $\hat J $ is  a  matrix  with  all elements equal to unity: $J^{nm}=1$.
The bare propagator thus becomes a matrix in the replica space:
\be
\hat G_\mathbf k^0  = \frac{T \hat \varkappa^{-1}}{ k^4} =\frac{T  }{ \varkappa k^4}\left( 1+  f  \hat J \right) ,
\label{G0}
\ee
where
\be
f=\frac{b \varkappa}{ T }
\ee
is the dimensionless parameter   characterising  a ratio between the bare disorder, $b,$ and the bare dynamical fluctuations, $T/\kappa.$

The polarization operator also becomes a replica-space matrix. In the one-loop order (i.e., within RPA), its elements read
\begin{eqnarray}
\label{Pinm}
  \Pi_\mathbf q^{nm} &=& \frac{1}{D^2-1} \int (dk) k_\perp^4 G_\mathbf k^{0,nm} G^{0,nm}_{\mathbf q-\mathbf k} \\
   &=& \frac{T^2}{D^2-1} \int (dk) k_\perp^4
   \frac{(\hat\varkappa^{-1})^{nm}}{k^4}\frac{(\hat\varkappa^{-1})^{nm}}{|\mathbf q-\mathbf k|^4} .
 \nonumber
\end{eqnarray}
Using the property $(\delta^{nm}+f J^{nm})^2 =(1+2f)\delta^{nm} +f^2 J^{nm}$, we find
\be
 \hat \Pi_\mathbf q =
 A_D \frac{T^2 }{\varkappa ^2 q^{4-D}} \left(1+2f +f^2 \hat J\right).
  \label{Pi0nm}
 \ee
Substituting Eq.~(\ref{Pi0nm}) into Eqs.~(\ref{N}) and (\ref{M}),
we obtain the screened couplings $\hat {\cal N}_\mathbf q$ and $\hat {\cal M}_\mathbf q$
as matrices in the replica space.
The
Ginzburg scale $q^*$ is now affected by the disorder that enters the polarization operator (\ref{Pi0nm}):
\begin{equation}
q^* \sim\left[\frac{\mu T (1+2f)}{\varkappa^2}\right]^{1/(4-D)}.
\label{q*dis}
\end{equation}
It is worth noting that for strong disorder or low temperatures ($f\gg 1$) $q^*\sim (\mu b/\kappa)^{1/(4-D)}$
is independent of temperature, while for weak disorder ($f\ll 1$) we recover Eq.~(\ref{q*}), $q^*\propto T^{1/2}$.

\begin{figure}[t]
%\center
\centerline{\includegraphics[width=0.4\textwidth]{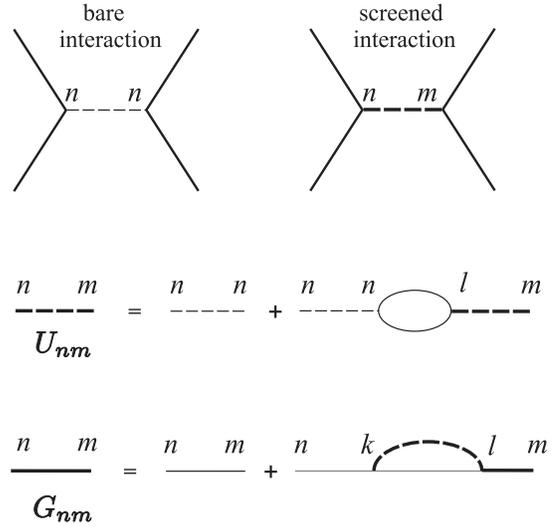} }
\caption{
 Replica-index structure of the $h^4-$interaction  and of the one-loop diagrams  for a  disordered  membrane.
}
\label{Fig3}
\end{figure}

Let us now  calculate the self-energy determining the renormalization of the bending rigidity.
In similarity with Eq.~\eqref{NM}, the effective interaction  $\hat U$ with matrix elements
$U_{nm}$ (see Fig.~\ref{Fig3}) is expressed for $q \ll q^*$  in terms of the polarization operator,
\BEA
 \hat  U  &=& \frac{D\hat \Pi_\mathbf q^{-1}}{2 (D+1)}   \label{NMnm} \\
 &=& \frac{D}{2 (D+1)} \frac{\varkappa^2 q^{4-D}}{T^2A_D}\frac{1+2f +f^2 N-f^2 \hat J}{(1+2f)(1+2f +f^2N)} .
\nonumber
\EEA
The replica generalization  of Eq.~\eqref{Sigma} for self-energy  thus takes the form
\be
\Sigma_\mathbf k^{nm}=\frac{2T}{d_c}\int (dq) k_\perp^4  \frac{D \left(\Pi_\mathbf q^{-1}\right)^{nm}}{2 (D+1)}G_{\mathbf k-\mathbf q}^{0,nm}.
\label{Sigmanm}
\ee
Substituting here the Green function, Eq.~\eqref{G0},   using the property
\BEA
&&\left[ {1+2f +f^2 N -f^2 \hat J}\right]^{nm} (1+f \hat J)^{nm}
\nonumber
\\
\nonumber
&&=
\left[{1+3f +f^2 (N+1)+f^3 N -f^3 \hat J}\right]^{nm},
\EEA
and taking the limit $N\to 0$, we finally arrive at the matrix equation which governs the renormalization of  $\hat \varkappa$ in the presence of disorder:
\be
\frac{d\hat \varkappa }{d\Lambda}= \eta \varkappa \frac{1+3f+f^2 - f^3 \hat J}{(1+2f)^2},
\label{dkappa1}
\ee
 where $\eta$ is given by  Eq.~\eqref{eta0}.
Substituting Eq.~\eqref{kappa-matrix} into l.h.s. of Eq.~\eqref{dkappa1}, and separating terms proportional to $\hat J$ from scalar ones, we find  two equations describing renormalization of the bending rigidity and of the disorder strength:
  \begin{eqnarray}
     \frac{d \varkappa }{d\Lambda}&=& \eta \varkappa \frac{1+3f+f^2}{(1+2f)^2}, \label{dkappa2}
      \\
     \frac{d (\varkappa f) }{d\Lambda}&=& \eta \varkappa  \frac{f^3 }{(1+2f)^2}.\label{dkappa3}
  \end{eqnarray}

%%%%%%%%%%%%%%%%%%%%%%%%%%
%%%%%%%%%%%%%%%%%%%%%%%%%%

Equations \eqref{RGxixi}, \eqref{dkappa2}, \eqref{dkappa3} represent the full set of RG equations which  determine the behavior of the system for $q\ll q^*$
[the Ginzburg scale $q^*$ is defined here with the bare disorder strength $f_0$ in Eq.~(\ref{q*dis})].
To analyse the fixed points and the scaling flow, we rewrite them in terms of appropriate dimensionless couplings. To this end,
we introduce a rescaled    bending rigidity,
\be
\tilde{\varkappa}=\varkappa \tilde{\xi}^2= \varkappa {\xi}^2 k^{2-D}.
\label{kappa-tilde}
\ee
The meaning of Eq.~\ref{kappa-tilde} is twofold. First, the ratio $\tilde\varkappa/T$ is made dimensionless (also for a dimensionality deviating from $D=2$). Second, it takes into account the ``compression'' of the membrane controlled by the factor $\xi$ in Eq.~(\ref{xi-new1}), which can be viewed as an analog of the order-parameter field renormalization in the conventional $\sigma$-model RG.
Expressing further the flow of $f$ from  Eqs.~\eqref{dkappa2} and \eqref{dkappa3}, we cast the RG equation into the following form:
   \begin{eqnarray}
\frac{d {\xi}^2}{ d\Lambda} &=&  -    \frac{\tilde {d}_c}{4\pi}\frac{T}{\tilde{\varkappa}} (1+f) {\xi}^2,
\label{RG-final-xi}
          \\
     \frac{d \tilde{\varkappa} }{d\Lambda}&=&  \tilde{\varkappa} \left [ D-2 + \eta \frac{1+3f+f^2}{(1+2f)^2} \right] -\frac{\tilde{d_c}  T }{4\pi} (1+f), \label{RG-final-kappa}\\
       \frac{d  f }{d\Lambda}&=& - \eta  \frac{f(1+3f) }{(1+2f)^2}.\label{RG-final-f}
              \end{eqnarray}

It is worth discussing a structure of the derived  RG equations. Equations (\ref{RG-final-kappa}) and (\ref{RG-final-f}) describe evolution of two dimensionless couplings of the theory: the inverse dimensionless bending rigidity $T/\tilde\kappa$ and the dimensionless disorder $\tilde b = fT/\tilde\kappa$.
The solution of one-loop equations is simplified by the fact that the equation for $f$ (ratio of disorder to the inverse bending rigidity),  Eq.~(\ref{RG-final-f}), decouples  and thus can be straightforwardly integrated. After this, Eq.~(\ref{RG-final-kappa}) for $\tilde\varkappa$ can be solved.
Finally,  Eq.~\eqref{RG-final-xi} is analogous to the wave-function renormalization equation of the $\sigma$-model RG. It is a ``slave" equation which can be solved at the last stage.

We emphasize that the set of the RG equations \eqref{RG-final-xi}, (\ref{RG-final-kappa}), and (\ref{RG-final-f})
strongly differs from the RG equations~\cite{disorders,disorder-imp,disorders-Morse-Grest} previously derived in
dimensionality $D=4-\epsilon$. Apart from the dimensionality (that allows us to employ the RG approach to physical $D=2$ graphene
samples), the crucial difference is in the role of screening near two dimensions. In our case, the screening of anharmonic couplings
eliminates the separate renormalization of elastic constants $\lambda$ and $\mu$; they now only determine the starting point of the RG, $q^*$,
Eq.~(\ref{q*dis}). For the same reason, only the out-of-plane (curvature) short-range disorder survives the renormalization,
see discussion of the in-plane disorder in Sec.~\ref{subs:in-plane} below.

One can further simplify Eqs. ~\eqref{RG-final-xi}  and \eqref{RG-final-kappa} by introducing a dimensionless coupling
 \be
              K= \frac{4\pi \eta \tilde{\varkappa} }{ \tilde {d}_c T (1+f)}
              \label{KK}
              \ee
[for $D=2$ this transformation takes the form
$K= ({8\pi}/{d_c^2 T}) {\tilde{\varkappa}}/{(1+f)}$].
We also change the variable $\Lambda$ to
              \be
              z=\eta \Lambda,
              \ee
              thus arriving at the following  RG equations:
              \begin{eqnarray}
\frac{d {\xi}^2}{ dz} &=&    - \frac{{\xi}^2 }{K},
\label{RG-final-xi1}
\\
\frac{d K }{dz}&=&   K\left [ \epsilon_2 +  \frac{1+5f+7f^2+f^3}{(1+f)(1+2f)^2} \right] -1,
             \label{RG-final-K}\\
              \frac{d  f }{dz}&=& -   \frac{f(1+3f) }{(1+2f)^2}.\label{RG-final-f1}
              \end{eqnarray}
Here
\be
\epsilon_2=\frac{D-2}{\eta}  \simeq \frac{(D-2)d_c}{2}.
\ee

Before proceeding to a detailed analysis of these RG equations, we briefly discuss the effect of in-plane disorder.

\subsection{In-plane disorder}
\label{subs:in-plane}
In most of this paper, we explore the problem in the presence of  out-of-plane (``random curvature'') disorder. In this subsection we briefly analyse the in-plane disorder and show that, with an exception of the case of topological defects (disclinations), it can be safely neglected.
We restrict ourselves  to the case $D=2$.  Let us consider an impurity-induced isotropic  in-plane disorder  that leads\cite{disorder-imp} to the following modification of  the  free energy \eqref{F}:
\BEA
\label{FF}
F&=& \frac{\varkappa}{2}\int (dk) k^4 |\mathbf h_\mathbf k|^2 \\
&+&\frac{\cal{N}}{4d_c}  \int (dq)\left| \int (dk) \frac{(\mathbf k\times \mathbf q)^2}{q^2} \mathbf h_{\mathbf k+\mathbf q}\mathbf h_{-\mathbf k}  +c_\mathbf q \right|^2.
\nonumber
\EEA
Here $c_\mathbf q$ is the Fourier  transform of the random quenched field $c(\mathbf x)$ with the Gaussian distribution
$$P[c(\mathbf x)] \propto \exp\left(  -\frac{1}{2\sigma} \int d\mathbf x c^2(\mathbf x)\right), $$
and $\sigma$ is the effective disorder strength.
Replicating  the field $\mathbf h$ and averaging over disorder,  we  find that  the bare interaction $U_{nm}^0$   (see Fig.~\ref{Fig3}) acquires now off-diagonal elements in the replica space:
\be
U_{nm}^0= {\cal N}\delta_{nm} -  {\cal N}^2 \sigma \hat J.
\label{Unm}
\ee
   Next, we find the screened interaction by solving the equation $\hat U=\hat U_0 - \hat U_0  \Pi \hat U$   (see Fig.~\ref{Fig3}), which yields
   \be
   \hat{U}= (1+\hat U_0  \Pi)^{-1}\hat U_0=(1+\Pi^{-1} \hat U_0^{-1} )^{-1} \Pi^{-1}.
   \label{U}
   \ee
    For $q\ll q^*$, the polarization operator grows as  $1/q^2$, so that  one can neglect the term $\Pi^{-1} \hat U_0^{-1} \sim q^2$. As a result,   both the elastic constant ${\cal N}$ and the  disorder strength $\sigma$ drop out from  the effective interaction:
    \be
    \hat U \to   \Pi^{-1},    \,\,\,\,\,\, \text{for} \,\,\, q \ll q^*.
        \label{Ueff}
    \ee
     We thus conclude that, in the critical region $q \ll q^*$,  the in-plane random potential becomes irrelevant and  does not affect critical behavior of  $\varkappa.$

     This conclusion should be contrasted with that of Ref.~ \onlinecite{RLD} where it was argued  that an arbitrarily  weak in-plane impurity disorder destroys the flat phase. Specifically, it was found that the membrane is destroyed   for  $\sigma >\sigma_c^{(0)},$ where $1/\sigma_c^{(0)}$ is strongly divergent in the infrared limit: $1/\sigma_c^{(0)} \propto 1/q_{\rm  IR}^2. $  Here,  $q_{\rm  IR}$ is the infrared cutoff momentum (inverse system size, $q_{\rm IR} \sim 1/L$).
         One can demonstrate  that in fact  screening cures this divergency  (cf. Ref.~\onlinecite{RLD1})  and  leads to a replacement of
     $q_{\rm  IR}$  with $q^*$, thus yielding a  finite value $\sigma_c \propto T/\mu$.
      A not too strong in-plane disorder, $\sigma_0 < \sigma_c$   will thus simply lead to ultraviolet corrections to the bare parameters (out-of-plane disorder $b_0$  and bending rigidity $\varkappa_0$) of our theory.
    Specifically,   a natural expectation is that the in-plane disorder would enhance $b_0$ and reduce $\varkappa_0$.
      In other words, the in-plane disorder is RG-irrelevant at length scales $\gtrsim 1/q_*$ but its contribution on shorter  (atomic) scales  can affect the starting point of RG.

The above result on irrelevance of the in-plane disorder was based on an assumption of a finite-range disorder $c(\bf{x})$, which means that the disorder strength $\sigma_\mathbf q$ remains finite in the limit $\mathbf q\to 0$. It is easy to see that this assumption can be in fact weakened:  it is sufficient to require that the singularity of $\sigma_\mathbf q$ at $\mathbf q\to 0$ (if present) is not stronger than $1/q^{2-\alpha}$ with $\alpha>0$. This corresponds to spatial correlations with a power-law decay, $\langle c(0)c(\mathbf x)\rangle \propto x^{-\alpha}$.
In this case,   inverting Eq.~\eqref{Unm} and taking the limit $N\to0,$  we find
 \be
 \hat U_0^{-1} (N\to 0)= \frac{1}{{\cal N}} +\sigma_q \hat J.
 \ee
 Since $\sigma_q \Pi^{-1} \to  q^\alpha $  for $q\to 0$ ,  we conclude that   $\hat U_0^{-1} \Pi^{-1}$ vanishes at small $q$, so that the derivation of  \eqref{Ueff} retains its validity.
Only for $\alpha \to 0$ (which corresponds to logarithmic real-space correlations) we get $\sigma_q \Pi^{-1} \to  \rm{const} $  for $q\to 0$.
Such a disorder is RG-marginal and thus will influence the flow of other couplings. Physically, the $1/q^2$ in-plane disorder corresponds to random topological defects---disclinations (see Ref.~\onlinecite{RLD1} for a discussion of different types of long-range disorder
and their treatment within the SCSA). A  detailed analysis of this type of in-plane disorder is outside of the scope of this
 paper and will be presented elsewhere.

\section{Analysis of RG equations: Crumpling transition }

In this section we  analyze  the RG  equations derived above  for clean and disordered cases.

\subsection{Clean membrane}

In the absence of disorder ($f=0$) and for $D=2$, Eqs.~\eqref{RG-final-xi1} and \eqref{RG-final-K}  take the form
\begin{eqnarray}
\frac{d {\xi}^2}{ dz} &=&  -\frac{ {\xi}^2}{ K},
\label{RG-clean-xi}
\\
\frac{d K }{dz}&=&    K  -1.
             \label{RG-clean-K}
              \end{eqnarray}
             Here we have taken into account that $\tilde {\xi}=\xi$  for $D=2.$   The initial condition for Eq.~\eqref{RG-clean-xi}  is \cite{foot-xi}
             \be
             \xi_0=1, \qquad \text{for}~ z=0.
             \ee
From Eqs.~\eqref{RG-clean-xi} and \eqref{RG-clean-K} we conclude that  there exists an unstable fixed point
\be
K_{\rm cr}= 1
\label{cr}
\ee
or, equivalently,
\be
 {\varkappa}_{\rm cr}  = \frac{d_c^2 T}{8\pi}.
 \label{cr1}
\ee
 Indeed, assuming that the starting value of  the bending rigidity, $\varkappa_0=  \tilde {\varkappa} _{z=0}= \varkappa_{z=0},$ exceeds the critical value, $\varkappa_0 >\varkappa_{\rm cr} $,  we find from  Eqs.~\eqref{RG-clean-xi} and \eqref{RG-clean-K}
\be
\xi^2= \frac{\varkappa_{\rm cr} e^{-z} + \varkappa_{0}-\varkappa_{\rm cr} }{\varkappa_{0}},
\label{xiz}
\ee
and, consequently,
\be
\xi^2_{z=\infty}=\frac{\varkappa_0-\varkappa_{\rm cr}}{\varkappa_0}.
\label{xiinf}
\ee
Therefore, above the critical point membrane remains in the flat phase in the course of renormalization. On the other hand, one can easily check that  below the critical point    (for  $\varkappa_0 <\varkappa_{\rm cr} $), the membrane shrinks to the crumpled phase, $\xi=0,$ at a finite scale
\be
L \sim q^{-1}= \frac{1}{q^*} \left ( \frac{\varkappa_{\rm cr}}{\varkappa_{\rm cr}- \varkappa_0 }\right)^{d_c/2}.
\ee
Hence, the fixed point \eqref{cr1} separates the crumpled and flat phases.\cite{david1}

For a membrane dimensionality $D$ slightly deviating from $2$,  we get instead of \eqref{RG-clean-K}
\be
\frac{d K }{dz}=   K\left (\epsilon_2 + 1  \right) -1.
             \label{RG-clean-K1}
\ee
Equation \eqref{RG-clean-K1} implies that  the lower critical dimension for crumpling transition can be found from the condition $\epsilon_2=-1$,  which yields
\be
D_{\rm cr}=2- \frac{2}{d_c},
\ee
in agreement with previous studies. \cite{lower-cr-D2}.

Exactly  at the transition point, when $K=K_{\rm cr}=1/(1+\epsilon_2),$  the stretching factor $\xi$ decays with $L$ according to  a power law:
\be
\xi \propto \frac{1}{L^\tau},
\label{xixi}
\ee
where $\tau = (D-2+\eta)/2$. In other words, the extension of the membrane $R$ in the embedding space scales with its ``intrinsic'' length $L$  as $R=L^{1-\tau}$.   The exponent $\tau$ determines thus the fractal (Hausdorff) dimension $D_H$ of the membrane at criticality (defined by the relation $R^{D_H} \sim L^D$), yielding $D_H = D/(1-\tau)= 2D/(4-D-\eta).$

Let us discuss in more detail geometric properties of the membrane, which are determined by the behavior of $\xi.$  To this end,  let us consider  two points $\mathbf r_1$ and  $\mathbf r_2$ on the membrane and the corresponding points $\mathbf x_1$ and  $\mathbf x_2$ in the reference plane. According to Eq.~ \eqref{xi-new1} and to the RG procedure, we have a scaling relation
\be
 \langle |\mathbf r_1 -\mathbf r_2|^2  \rangle  \sim \xi_{|\mathbf x_1-\mathbf x_2 |} ^2 (\mathbf x_1-\mathbf x_2)^2,
 \label{xr}
 \ee
where $\xi_x$ is the value of the running renormalization-parameter $\xi$ at the RG spatial scale $x$.  When $\xi_x$ drops down to a value substantially smaller than unity (say, to $1/2$), the embedding-space distance $|\mathbf r_1 -\mathbf r_2|$ between the points become essentially smaller than the intrinsic distance $|\mathbf x_1-\mathbf x_2|$. This indicates that a membrane of such size starts to show ``folding'' (i.e., strong spatial variation of the normal vector to the membrane in the embedding space).

Deep in the flat phase,  the renormalization of  $\xi$ at the whole interval of RG scales, $1/q^*<L<\infty$, is relatively weak.  This implies that the membrane does not fold. In other words, although the surface of a membrane is not exactly flat due to dynamical fluctuations (and also due to ripples in the disordered case, as discussed below), the spatial variation of the normal vector remains relatively small.

On the other hand, when  $\varkappa_0$  approaches the critical value,   Eq.~\eqref{cr1} (i.e., when $\varkappa_0-\varkappa_{\rm cr } \ll  \varkappa_{\rm cr })$,  the membrane shows fractal folding at the broad interval of length scales.  Specifically, as seen from Eq,~\eqref{xiz}, this folding becomes strong for  $z \sim 1,$ i.e. at the length scale $L \sim L_1,$  where
\be
L_1 \sim \frac{1}{q^*} e^{ 1/\eta}.
\label{L1-clean}
\ee
At a much larger scale,
   \be
  L_2 \sim \frac{1}{q^*} \left( \frac {\varkappa_{\rm cr } }{\varkappa_0-\varkappa_{\rm cr } }\right) ^{1/\eta},
\label{L2-clean}
   \ee
    $\exp(-z)$  becomes  on the order of  $(\varkappa_0-\varkappa_{\rm cr } )/ \varkappa_{\rm cr }  $ and   $\xi $ saturates.  Thus, the membrane has a fractal geometry in the interval $L_1\ll L \ll L_2$.   At larger scales, $L \gg L_2$ the membrane flattens. Exactly at the transition, $L_2$ diverges and the membrane remains fractal at arbitrarily large scales.

\subsection{Disordered  membrane}

\subsubsection{RG flow in disordered case }

\begin{figure}[t]
%\center
\centerline{\includegraphics[width=0.35\textwidth]{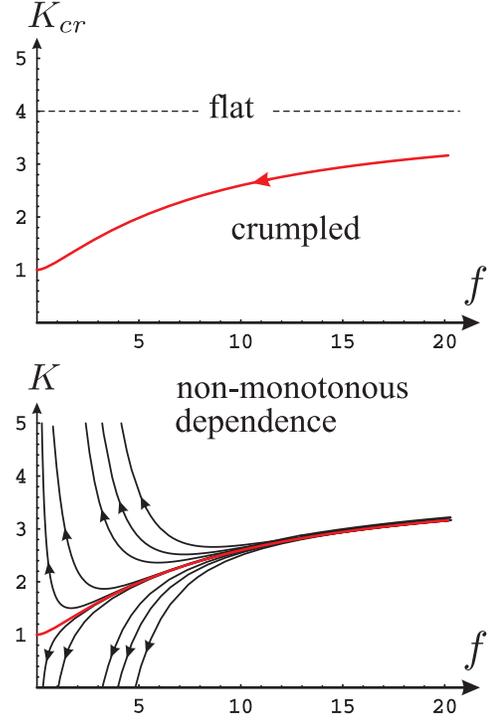} }
\caption{
{\it Upper panel:} Critical curve ($\delta=0$)  separating the crumpled and flat phases in the $K$ - $f$ plane.
{\it Lower panel:}  RG flow lines  in the $K$ - $f$ plane for a disordered membrane for different values of $\delta$  ($\delta$= -200, -100, -50, -3, -0.4, 0, 1.2, 7.5, 50, 100, 200 from bottom to top).
}
\label{Fig4}
\end{figure}

 Let us now consider the disordered case. From now on and till the end of the paper, we will assume that   $D=2$ and, consequently, $\epsilon_2=0$.    We see from Eq.~\eqref{RG-final-f1}  that $f$ monotonously decreases in course of renormalization.  Dividing Eq.~\eqref{RG-final-K}  by Eq.~\eqref{RG-final-f1}, we find an equation that determines variation of $K$ with $f$. Its solution  yields the function $K(f)$  which  can be written in the  following form:
 \be
 K(f)=K_{\rm cr}(f)+ \frac{(3f+1)^{1/9}e^{-f/3}}{f(1+f)} \delta.
 \label{Kf}
 \ee
  Here $K_{\rm cr}(f)$ is the critical curve in the $(K,f)$ plane which separates  the crumpled and flat phases (see Fig.~\ref{Fig4}):
   \be
 K_{\rm cr}(f)=\frac{(3f+1)^{1/9}e^{-f/3}}{f(1+f)}\int\limits_0^f dy \frac{(1+2y)^2(1+y) e^{y/3}}{(3y+1)^{10/9}}.
 \label{Kcrf}
 \ee
The phase boundary has the following asymptotic behaviour at small and large $f$:
 \be
 K_{\rm cr}\simeq \left \{ \begin{array}{cc} \displaystyle
                  1+{f^2}/{3} , & \qquad \text{for}~ f\to0;\\ \displaystyle
                 4 \left(1-{5}/{f} \right) , & \qquad \text{for}~f\to\infty.
               \end{array} \right.
 \label{asympt}
 \ee
 For $f=0$ we recover  Eq.~\eqref{cr} as expected.
The parameter $\delta$ controlling the deviation of the RG flow line from  the critical one depends on initial values of  $f$ and $K$ in the following way:
 \be
 \delta=\frac{f_0(1+f_0)}{(3f_0+1)^{1/9}} e^{f_0/3} [K_0-K_{\rm cr}(f_0)].
\label{delta}
 \ee
 The flat phase corresponds to $\delta>0$,   while  the crumpled phase to $\delta <0$.

%%%%%%%%%%%%%%%%%%%%%%%%
%%%%%%%%%%%%%%%%%%%%%%%%%

The RG flow in the ($K , f$)  plane is illustrated in Fig.~(\ref{Fig4}).  One observes that  in the flat phase the evolution of $K$ is non-monotonous.    For $\delta \ll 1$  the minimum of $K$ lies in the region of small $f$, where $K\approx 1+f^2/3+\delta/f$.  Minimization yields the position of the minimum:
\be
 f_{\rm min}\approx (3\delta/2)^{1/3}, ~K_{\rm min} \approx 1+(3\delta/2)^{2/3} ,~\text{for }~\delta \ll 1 .
\label{deltasmall}
\ee
In the opposite case, $\delta \gg 1$,   the minimum corresponds to large $f$, where $K \approx 4-20/f +9\delta \exp(-f/3)/(3f)^{17/9}$.  We find with logarithmical precision
\be
f_{\rm min}\approx 3 \ln (\delta), ~ K_{\rm min} \approx4-20/3\ln(\delta),  ~\text{for }~\delta \gg 1 .
\label{deltalarge}
\ee

\subsubsection{Disorder-induced crumpling }

Let us fix initial value of $\varkappa_0=  \tilde{\varkappa}_{z=0} =\varkappa_{z=0}$ and consider what happens with increasing disorder strength $b$.  Since $\xi_0=1$, we have the following initial values of $K$ and $f$:
\be
K_0=  \frac{\varkappa_0}{\varkappa_{\rm cr} (1+f_0)}, \qquad f_0 =\frac{b_0 \varkappa_0}{T},
\ee
where   $\varkappa_{\rm cr}$ is given by Eq.~\eqref{cr1}.  The crumpling occurs when $K_0$ becomes smaller than  $K_{\rm cr}(f_0).$ Hence, the critical curve in  the $(\varkappa_0,b_0)$ plane   is implicitly determined by equations ${\varkappa_0}={\varkappa_{\rm cr} (1+f_0)}K_{\rm cr}(f_0)$ and
$b_0=  f_0 T/ \varkappa_0= (8\pi/d^2) f_0/ (1+f_0)K_{\rm cr}(f_0)$, with $f_0$ varying in the interval $(0,\infty).$   The dependence $b_0 (\varkappa_0)$ found from these equations is  plotted in Fig.~(\ref{Fig5}).
The asymptotic behaviour of the critical line reads
\be
b_0 =  \frac{2\pi}{d_c^2} \left \{\begin{array}{c}
        4(\varkappa_0-\varkappa_{\rm cr})/\varkappa_{\rm cr}, \qquad \text{for}~   \varkappa_0 \to \varkappa_{\rm cr};\\
       1+19\varkappa_{\rm cr}/16\varkappa_0, \qquad \text{for} ~  \varkappa_0 \to \infty.
     \end{array} \right.
\ee
 We notice  that the dependence is non-monotonous. In other words, there exists an optimal value of  bare  bending rigidity, $\varkappa_0\approx 4.2\varkappa_{\rm cr},$  for which the membrane is most robust to disorder-induced crumpling.  It is also worth stressing that  the critical disorder saturates in the limit of infinite bending rigidity at the universal value $b_0 =2\pi/d_c^2.$

 \section{Rippling and folding of membrane in the flat phase}

We have thus established the phase diagram of the crumpling transition in the plane of initial parameters, i.e., bending rigidity $\varkappa_0$ and disorder $b_0$. An important question that remains to be explored is the evolution of physical observables with the length scale. This will be done in the present section. We will, in particular, show that the disorder strength decreases beyond a certain length scale $L_r.$

%%%%%%%%%%%%%%%%%%%%%%%%
\begin{figure}[t]
\centerline{\includegraphics[width=0.35\textwidth]{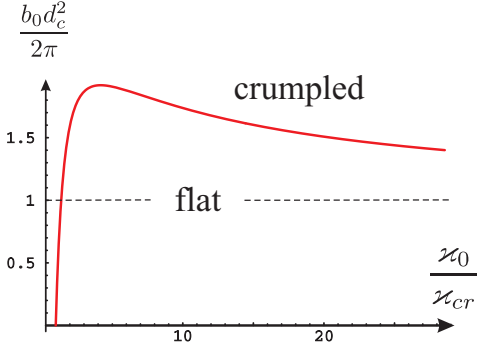} }
\caption{
Crumpling-transition phase diagram in the plane of initial parameters (bending rigidity $\varkappa_0$ and disorder $b_0$).
}
\label{Fig5}
\end{figure}
%%%%%%%%%%%%%%%%%%%%%%%%%
%%%%%%%%%%%%%%%%%%%%%%%%%%%%%%%
\begin{figure}[t]
\centerline{\includegraphics[width=0.45\textwidth]{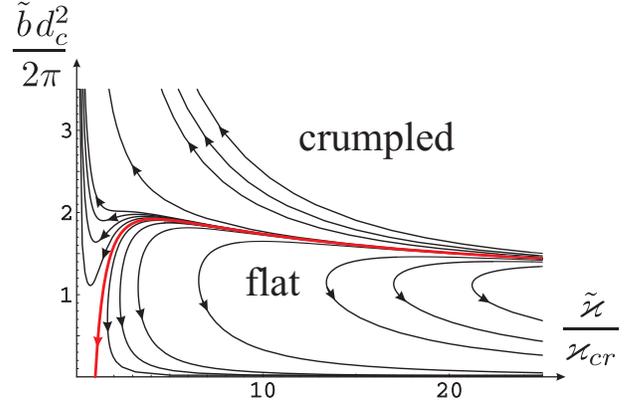} }
\caption{
RG flow  in the  $(\tilde{\varkappa}, \tilde{b})-$plane for  $\delta$ = -200, -100, -50, -3, -0.4, -0.3, -0.2, -0.1, 0, 0.1, 0.3, 1.2, 7.5, 50, 100, 200,  increasing from top to bottom.
}
\label{Fig6}
\end{figure}
%%%%%%%%%%%%%%%%%%%%%%%%%%%%%%%

After rescaling  the   disorder strength
\be
\tilde{b}=\frac{b}{\xi^2},
\ee
one finds  that the RG equations in the $(\tilde{\varkappa}, \tilde{b})-$plane are implicitly determined by the following equations
\BEA
\frac{\tilde{\varkappa}}{ \varkappa_{\rm cr}}&=& (1+f) K(f),\label{kappa-fin}
\\
\frac{\tilde{b} d_c^2}{2\pi}&=& \frac{4f}{(1+f) K(f)}, \label{b-fin}
\EEA
where $K(f) $ is given by Eq.~\eqref{Kf} and  $f$ varies in the interval $[0,\infty]$. For $\delta=0$ we reproduce the critical curve  $b_0(\varkappa_0)$ separating the crumpled and flat phases, see Fig.~\ref{Fig5}.  The RG flow lines are illustrated in Fig.~\ref{Fig6}.   We see that in the flat   phase    at the first stage of the renormalization the bending rigidity strongly decreases, while the disorder strength slightly increases.  This behavior indicates the tendency of the membrane to the disorder-induced crumpling. However, below the critical line the disorder strength is insufficient to destroy the membrane, so that $\tilde {\varkappa}$ eventually shows a minimum and then starts to grow, while the disorder  gets suppressed.   Close to the minimum of $\tilde {\varkappa}, $  the RG curve  is  approximately  vertical. This behavior  can be interpreted as screening of disorder by softened membrane.

The minimal value of  the bending rigidity is given by
\be
\tilde {\varkappa}_{\rm min} \simeq \varkappa_{\rm cr} \left\{
 \begin{array} {c}
1+2\sqrt{\delta}, ~\,\,\, \text{for} ~ \delta \to 0,
\\
12 \ln \delta, ~\,\,\,\text{for}~ \delta \to \infty.
\end{array} \right.
\ee
A particularly interesting behaviour
is predicted  slightly above the critical curve ($\delta<0$ with $|\delta| \ll 1$). In this case,  the disorder first increases, then  reaches a maximum, starts to decrease and, finally, after reaching a deep minimum, increases  again and goes to infinity.

The geometry of membrane is determined by the behavior of the stretching  factor $\xi,$ which  can be expressed in terms of $K(f)$  by using  Eqs.~\eqref{RG-final-xi1} and \eqref{RG-final-f1}:
\be
 \xi^2=  \exp\left[-  \int\limits_f^{f_0} dy\frac{(1+2y)^2}{y(1+3y) K(y)}\right].
\label{xif}
 \ee
The value of $\xi$ in the limit $L\to \infty$ is found from this equation by putting  $f=0:$ $\xi_{z=\infty}= \xi_{f=0}.$

\begin{figure}[t]
\centerline{\includegraphics[width=0.50\textwidth]{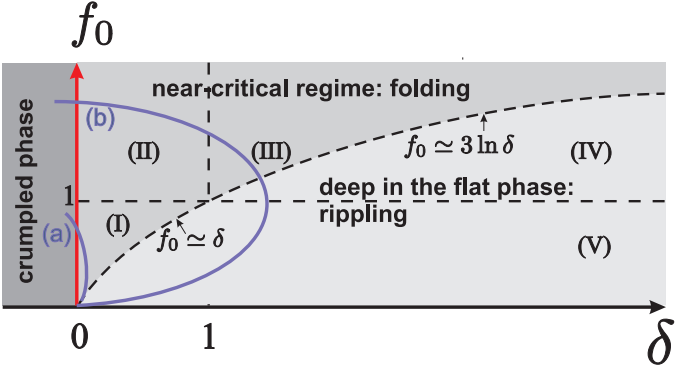} }
\caption{
Phase diagram in the $(\delta , f_0)$ plane. Critical curve (shown in red) separates flat and crumpled phases.  Clean case corresponds to horizontal axis ($f_0=0$).   Regions (I), (II), (III) correspond to a near-critical regime within the flat phase. In this part of the phase space, the membrane shows critical (fractal) folding at intermediate scales before flattening at larger scales.  Regions (IV) and (V) correspond to  a rippled membrane  deep in the flat phase.   Blue curves correspond to fixed values of bare bending rigidity $\varkappa_0$:  (a) $(\varkappa_0-\varkappa_{\rm{cr}})/\varkappa_{\rm{cr}} \ll 1;$   (b)$(\varkappa_0-\varkappa_{\rm{cr}})/\varkappa_{\rm{cr}} \gg 1.$   Bare disorder increases along these curves from the bottom to the top.}
\label{Fig7}
\end{figure}

Having in mind applications to  graphene,  let us now discuss   in more detail the scale dependence of  the  bending rigidity and disorder in the flat phase ($\delta>0$).  Similar to the clean case, the behavior of membrane is essentially different deep in the flat phase (when $\delta $ is large)  and  in the near-critical regime (when $\delta$ is small).    The regions corresponding to  different regimes are shown schematically  in Fig.~\ref{Fig7} on the $(\delta,f_0)$ plane.  Regions (I), (II), and (II) are sufficiently close to critical line, so that the membrane undergoes strong folding  in course of the  renormalization. Therefore, we term these regions near-critical.        In contrast,  for membranes with a starting point  within  regions  (IV) and  (V), the stretching parameter $\xi$ does  not change essentially  in course of renormalization (i.e., $\xi$ remains close to unity in the limit $L\to \infty$).   Such membranes do not fold and only show  small dynamical wrinkling  and   static rippling.

%%%%%%%%%%%%%%%%%%%%%%%%%%%%%%%%%%%%%%%%5

  We consider separately  the cases of small and large deviations from the critical  transition line.
%\vspace{5mm}

\subsection {Close to critical line [regions (I), (II), and (III) in Fig.~\ref{Fig7}].} \label{near}

When the system is in the flat phase but not far from the transition (i.e,  $\delta > 0$ but  $\delta \ll 1,$  regions (I) and (II) in Fig.~\ref{Fig7}),    the dependence   $\tilde {\varkappa} (f)$ has a minimum  at $f \simeq \sqrt{\delta}$.  The minimal value is given by  $(\tilde{\varkappa}/\varkappa_{\rm cr})_{\rm min} \simeq 1+2\sqrt{\delta}$ and therefore is very close to the critical value for the clean membrane. The disorder strength shows a maximum at a much larger values of $f$, namely,   $f \simeq 2.$ The maximal value of disorder is given by $\tilde{b}^{\rm max} d_c^2/2\pi \simeq  1.9$.

For strong disorder, when  the   starting values of $f$ is large, $f_0 \gg 1$ [region (II)], the overall RG evolution is particularly rich.
At the initial stage of renormalization (i.e., as long as $f$ remains large),  we find  form Eq.~\eqref{RG-final-f1} that $f$ changes linearly  with $z$:
\be
f(z)=f_0- \frac{3 z}{4}, \qquad \text{for}\, f\gg 1.
\label{fz}
\ee
As follows from Eq.~\eqref{b-fin},  at such   spatial  scales the disorder slowly increases,
\be
\frac{\tilde{b}(z)d_c^2}{2\pi}\simeq 1+\frac{4}{f(z)}.
\label{b-large}
\ee
 Here we have taken into account that $\delta \ll 1$, replaced  $K$ with $K_{\rm cr}$ in the denominator of Eq.~\eqref{b-fin}, and used the large-$f$ asymptotic of the function $K_{\rm cr}$  [see Eq.~\eqref{asympt}].
 The bending rigidity decreases  linearly with $z$:
 \be
 \frac{\tilde{\varkappa} }{\varkappa_{\rm cr}} \approx  4 {f(z)}
 \label{kappa-large}
 \ee

 The scaling of disorder changes when $f$ becomes of the order of unity.  This happens for
  \be
  z_r \approx \frac{4f_0}{3} + { O}[\ln f_0]
  \ee
    As  seen from Eq.~\eqref{RG-final-f},  for  large scales $f$ decays exponentially with $z$, i.e.,   $f (z)\sim \exp(-z),$ so that the disorder starts to decrease as
\be
\frac{\tilde{b}(z)d_c^2}{2\pi}\approx \frac {4 f(z)}{K[f(z)]}.
\ee
Although the disorder strength starts to fall, the bending rigidity  is still strongly affected by disorder and continues to decrease  up to a value $\tilde{\varkappa}_{\rm min}$ where $f $ becomes    quite small (of the order of $\sqrt{\delta}$). Only  after this the bending rigidity  begins  to grow.

In the case when the starting value of  disorder is weaker,  $  \sqrt{\delta} \ll f_0\ll1$  [region (I)],     it does not show a maximum and  monotonously decreases  in course of the renormalization. However, as was mentioned in the previous paragraph, the disorder  strongly affects $ \tilde{\varkappa}$ as long as  $f>\sqrt{\delta}.$

The disorder becomes irrelevant for the evolution of bending rigidity for $f_0 \ll \sqrt{\delta}.$    In this case,  $f\approx f_0 \exp(-z)$  and, using Eqs.~\eqref{Kf},  \eqref{asympt}, \eqref{delta}, \eqref{kappa-fin}, and \eqref{b-fin},  we  arrive at the following scaling dependencies
\BEA
\tilde{\varkappa} &\approx& \varkappa_{\rm cr}+ (\varkappa_0-\varkappa_{\rm cr}) e^z, \label{kappa-fin1}
\\
\tilde{b}&\approx & \frac{b_0\varkappa_0 e^{-z}}{\varkappa_{\rm cr}+(\varkappa_0-\varkappa_{\rm cr})e^z}. \label{b-fin1}
\EEA
The disorder strength  decreases in course of the renormalization, first as $\exp(-z)$  and, at larger scales, when bending rigidity becomes large enough, as $\exp(-2z).$   A crossover between these two regimes happens at $f\sim \delta$, i.e. on the boundary separating the near-critical  and  flat phases.

Using Eq.~\eqref{xif}, one can easily find that in regions  (I) and (II)  the membrane folds in the course of the renormalization. In particular, in the region (II), we find  that at the early stage of renormalization (when  $f \gg 1$) the rescaling parameter $\xi$ is given by    $\xi^2 \approx \exp[ (f_0-f)/3 ] \approx \exp[-z/4]=  (q^*L)^{-\eta/4}  . $      Hence, first folding of membrane occurs at $z\approx 4.$ This yields the following estimate for  the length scale of folding:
\be
L_1\sim \frac{1}{q^*} e^{4/\eta}.
\label{L1-dirty}
\ee
 One can also find that $\xi$ saturates at an exponentially small value
 \be
 \xi_{z=\infty}^2 =\frac{\delta}{ f_0}e^{-f_0/3}
 \ee
for $L\sim L_2,$
where
\be
L_2 \sim \frac{1}{q^*} \left( \frac{f_0}{\delta}\right)^{1/\eta} e^{4f_0/3\eta}.
\label{L2-dirty}
\ee
It is worth stressing that $L_2$  coincides (up to a prefactor) with a length scale corresponding to maximum of disorder.

  In the region (I) disorder does not essentially affect the folding of the membrane, i.e., the dynamical fluctuations dominate.
 Therefore,  $L_1$ and $L_2$ are given by Eqs.~\eqref{L1-clean}  and \eqref{L2-clean} derived above for the clean case.

 Let us now discuss the region (III), where $  f_0 \gg  3\ln \delta \gg 1.$
 Using the asymptotic of $K(f)$ at large $f$ [see Eqs.~ \eqref{Kf} and \eqref{asympt}], we find that  the   functions  $\tilde{\varkappa} (f)$ and   $\tilde{b} (f)$  have, respectively, minimum and maximum at
$f_m \approx 3 \ln \delta.  $    The minimal value of rigidity is much larger than  the critical value,  $  {\tilde{\varkappa}_{\rm min} }/{\varkappa_{\rm cr}} \approx 12 \ln \delta, $ while the maximal value of disorder slightly exceeds  the universal value   $2\pi/d_c^2$; specifically,
$\tilde{b} d_c^2/2\pi -1 \sim 1/f_m. $

 In view of the assumption $f_0> f_m,$  the  maximum  in $\tilde {b}$  is reached at
\be
z_r =\frac{4}{3} (f_0 - 3 \ln \delta ) .
\ee
For $z \ll z_r$  the behavior of  $\tilde{b}$ and  $\tilde{\varkappa}$ can be well  described by  Eqs.~\eqref{b-large} and \eqref{kappa-large} with $f(z)$ given by Eq.~\eqref{fz}.  For
$z \gg z_r, $  one can neglect $K_{\rm cr}$ in expression \eqref{Kf}, so that  Eqs.~\eqref{kappa-fin}  and \eqref{b-fin} become
 \BEA
\frac{\tilde{\varkappa}}{ \varkappa_{\rm cr}}&=& \frac{\delta (3f+1)^{1/9}}{f} e^{-f/3},\label{kappa-fin11}
\\
\frac{\tilde{b} d_c^2}{2\pi}&=& \frac{4f^2}{\delta (3 f +1)^{1/9} } e^{f/3}. \label{b-fin11}
\EEA
 Since $f_m\gg1,$ there exists a large interval of $z,$ where these formulas can be used and at the same time $f\gg1$, so that Eq.~\eqref{fz} is applicable. In this interval we get, with exponential precision,
 \be \tilde{\varkappa} \propto e^{z/4},\qquad \tilde{b}\propto e^{-z/4}.  \label{kappa-b}\ee
Finally, for very large  $z,$ when $f$ becomes smaller than unity  and  decays as $\exp(-z),$ we find from Eqs.~\eqref{kappa-fin11} and \eqref{b-fin11}:
\be \tilde{\varkappa} \propto e^{z},\qquad \tilde{b} \propto e^{-2z}.   \label{kappa-b1}\ee
The latter regime   is well described  by Eqs.~\eqref{kappa-fin1} and \eqref{b-fin1} with $\varkappa_0-\varkappa_{\rm cr} \gg \varkappa_{\rm cr}.$

In analogy with the regions (I) and (II), we find that in the region (III) the membrane undergoes a critical folding in a broad range of length scales, $L_1<L<L_2$.  The value of $L_1$ is given by the same equation \eqref{L1-dirty} as in the region (II).  The stretching factor saturates at the value
\be
 \xi_{z=\infty}^2 =\delta e^{-f_0/3}
 \ee
 when the spatial scale $L$ becomes of the order of
  \be
L_2 \sim \frac{1}{q^*}  e^{4(f_0-3\ln\delta)/3\eta}.
\label{L2-dirty1}
\ee
 Again, up to a prefactor  this length coincides with the length $L_r$ where disorder has a maximum.

\subsection{Away from the critical line  [regions (IV) and (V) in Fig.~\ref{Fig7}].} \label{away}

The evolution of  bending rigidity and disorder for a membrane with initial couplings within the region (IV)  or  (V) coincides  with the intermediate (or, respectively, final)  stage of renormalization of $\tilde \varkappa$ and  $ \tilde b$  for the case of bare couplings in the region  (III).    Specifically,    both  $\tilde \varkappa$ and  $ \tilde b$ change monotonously with increasing  spatial scale: $\tilde \varkappa$ increases, while   $ \tilde b$ decreases. For the region (IV) there exists a large interval of length scales $L$ with  $ 1<f <3\ln \delta,$ where  Eqs.~\eqref{kappa-fin11}, \eqref{b-fin11},  and \eqref{kappa-b} apply.   With further increase of $L,$  we  enter the region $f<1,$  where scaling dependencies change to Eq.~\eqref{kappa-b1}.  For the region (V) the asymptotic formulas      \eqref{kappa-b1} are valid from the very beginning of the renormalization.

 Let us analyze the evolution of the stretching factor  $\xi$ in these two regions. For region (V)    the membrane stretching is fully controlled by dynamical fluctuations, while the disorder is irrelevant.   Hence, one can use Eqs.~\eqref{xiz} and \eqref{xiinf} derived for the clean case. One can easily check that within the region (V) $\varkappa_0- \varkappa_{\rm cr}  \gg   \varkappa_{\rm cr},$
so that $\xi$ does not change essentially  and, consequently, the membrane does not fold, remaining approximately flat at all scales.
The same statement is valid for region (IV) as well. Indeed, in this case, one can use the inequality $ 1<f <3\ln \delta$ and replace the  function $K(f) $ in Eq.~\eqref{xif}  with  $3^{1/9}\delta~ f^{-17/9}\exp(-f/3) \gg 1 $ [see Eq.~\eqref{Kf}].  Estimating then the integral in Eq.~\eqref{xif}, we find    that the renormalization of $\xi$ remains small.

This completes the analysis of the RG flow of the bending rigidity and the disorder for membranes with bare couplings in different regions of the flat phase. We have seen that the evolution can be very non-trivial, with several intermediate scaling regimes. It is worth emphasising, however, that the scaling behaviour at longest scales, $L\sim {q^{-1}} \to \infty$, is the same for the whole flat phase: the bending rigidity  increases, while the disorder gets suppressed according to universal power laws,
 \be
 \tilde\varkappa \propto  L^{\eta}, \qquad  \tilde b \propto L^{-2\eta} \qquad \text{for} \, L\to \infty,
 \ee
 with $\eta\simeq 2/d_c$ in the large-$d_c$ limit.

\subsection{Spatial scale of ripples }\label{spatial}

As follows from the above discussion,   the scaling dependence of both $\tilde{\varkappa}$ and $\tilde{b}$ in the flat phase are
especially non-trivial when the bare disorder is sufficiently strong. In particular, $\tilde{b}$ shows in this case a maximum  in the course of renormalization.
The spatial scale corresponding to the maximum can be easily found from  expression for $z_r$ and is given by
\be
L_r \simeq \frac{1}{q^*}  \left \{  \begin{array}{cc}
              e^{d_c f_0/6},  & \qquad \text{for}  \,\,\delta \ll 1, \ \ \ f_0\gg 1;\\
              e^{d_c (f_0-3\ln\delta)/6},  & \qquad \text{for}  \,\,\delta \gg 1, \ \ \ f_0> \ln\delta.
                         \end{array}\right.
            \label{lr}
\ee
For $L< L_r$ the disorder slowly increases with $z$, while  for $ L > L_r$  it decays exponentially with  $z$  (i.e., according to a  power law with respect to $L$).  In other words, the disorder is, roughly speaking,  ``switched off'' at $L>L_{r}.$
Hence, one can  interpret  $L_{r}$ as a characteristic scale of random static deformations---ripples. It is worth emphasising at this point that in the case of a nearly critical membrane such ``ripples'' are multiply folded (fractal) configurations.

%%%%%%%%%%%%%%%%%%%%%%%%%%%
\begin{figure}[t]
\centerline{\includegraphics[width=0.4\textwidth]{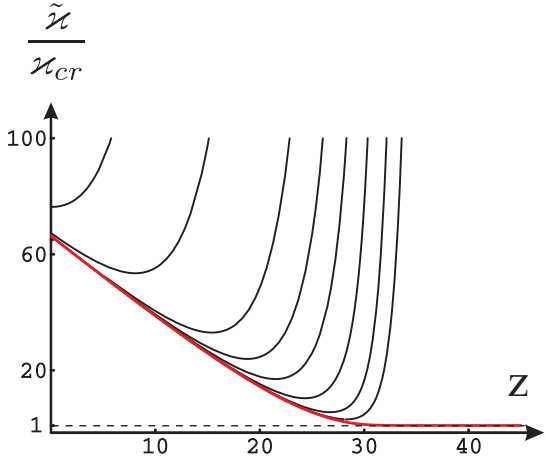} }
\caption{
Scale dependence of  the renormalized bending rigidity $\tilde{\varkappa}$     in the flat phase  for  $\delta$=1, 5, 25, 100, 300, 1000, 10000, 100000, increasing  from bottom to top. Critical curve ($\delta=0$) is shown in red.
}
\label{Fig8}
\end{figure}
%%%%%%%%%%%%%%%%%%%%%%%%%%%%

%%%%%%%%%%%%%%%%%%%%%%%%%%%%
\begin{figure}[h]
\centerline{\includegraphics[width=0.35\textwidth]{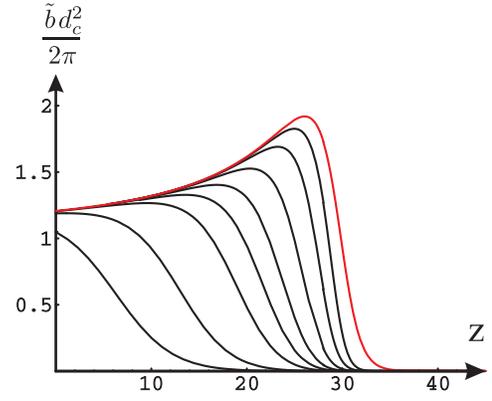} }
\caption{
Scale dependence of  the renormalized disorder $\tilde{ b}$
in the flat phase  for  fixed $f_0=20$ and for $\delta$=1, 5, 25, 100, 300, 1000, 10000, 100000, increasing  from top to bottom. Critical curve ($\delta=0$) is shown in red.
}
\label{Fig9}
\end{figure}
%%%%%%%%%%%%%%%%%%%%%%%%%%%%%%

 When the bare disorder is weaker [regions (IV) and (V) in Fig.~\ref{Fig7}], the ripples take a more conventional form
 of relatively small static deformations of a nearly flat surface. Indeed, in this case $\xi \approx 1$, so that the membrane does not fold.
For both regions,
the suppression of disorder  begins already on the Ginzburg length $1/q^*.$   The scaling of disorder at the initial stage of renormalization  is different for regions (IV) and (V):
\be
b\propto \left(q^*L \right)^{-\eta/4},
\ee
 for region (IV) and
\be
b\propto \left(q^*L \right)^{-2\eta},
\ee
for region (V). Hence,  the ripple  scale can be estimated as
\be
L_r\sim  \frac{1}{q^* }  \left \{\begin{array}{c}
                            e^{4/\eta}, \qquad \text{for} \,\,\, {\rm region\, (IV)},\\
                           e^{1/2\eta}, \qquad \text{for} \,\,\, {\rm region\, (V)}.
                         \end{array} \right.
\label{lr-ginzburg}
\ee
While  $\eta$ is small   ($\eta=2/d_c$) in the limit of high spatial dimensionality of the embedding space, it is a number of order unity, $\eta \simeq 0.7 \div 0.8$, for a physical 2D membrane (e.g., graphene) in a 3D space.
This yields  the characteristic scale for ripples of the order of Ginzburg length.~\cite{foot-Ginz}

%%%%%%%%%%%%%%%%%%%%%%%%%%%%%%
\begin{figure} [t]
\centerline{\includegraphics[width=0.35\textwidth]{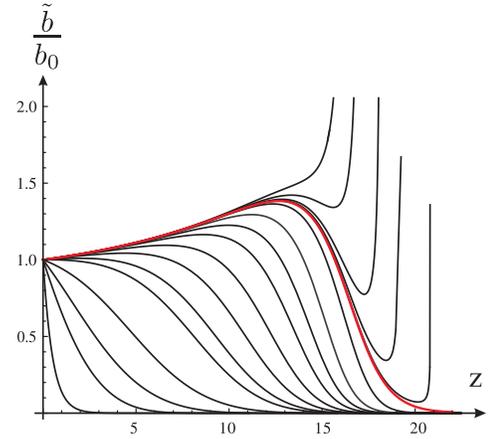} }
\caption{
Scale dependence of  the renormalized   disorder $\tilde{ b}  $  (measured in the units of its bare value $b_0$)  for $\varkappa_0/\varkappa_{\rm cr}=30 $
and   for  different value of  $b_0 d_c^2/2\pi$= 0.001, 0.5,  0.9, 1.2,  1.34, 1.36, 1.37525, 1.382,  1.385, 1.386, 1.38672, 1.3867555, 1.38676, 1.38677, 1.3868, 1.38685  increasing  from  bottom to top. Critical curve ($\delta=0, b_0=1.38675444319$) is shown in red.
}
\label{Fig10}
\end{figure}
%%%%%%%%%%%%%%%%%%%%%%%%%%%%%%

The functions $\tilde {\varkappa} (z) $ and   $\tilde {b} (z) $  are plotted in Figs. \ref{Fig8} and  \ref{Fig9} for $f_0=20$ and different  values of $\delta.$      These figures nicely illustrate the non-monotonous scale dependencies of the coupling constants,
with the bending rigidity $\tilde{\varkappa} (z)$  having a minimum and the disorder $\tilde{b}(z)$  showing a maximum, in full agreement with the above analytical results.

\begin{figure} [t]
%\center
\centerline{\includegraphics[width=0.4\textwidth]{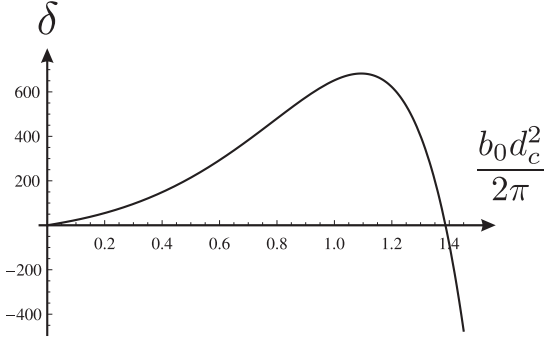} }
\caption{
Dependence of the parameter $\delta$ (which labels RG flow lines) on   the  bare disorder $b_0 d_c^2/2\pi$  for fixed bending rigidity, $\varkappa_0/\varkappa_{\rm cr}=30$. The sign change of $\delta$ corresponds to the crumpling transition.
}
\label{Fig11}
\end{figure}

 We also plot the dependence of disorder on the spatial scale for a fixed value  of $\varkappa_0 = 30 \varkappa_{\rm cr}$ and various values of the bare  disorder $b_0$  (see Fig.~\ref{Fig10}). In this case, starting points of the RG flow  lie on a vertical line in Fig.~\ref{Fig6}.  This plot serves as a nice illustration of the dependence of the characteristic scale $L_r$ on the bare disorder $b_0$. Indeed, it is seen, that for small $b_0$ the disorder drops quickly at the Ginzburg scale,  in agreement with Eq.~(\ref{lr-ginzburg}). With increasing $b_0$ the length $L_r$ increases. This effect becomes particularly strong when disorder becomes comparable to the critical one, i.e., the parameter $\delta$ approaches zero, as predicted by Eq.~(\ref{lr}). For the near-critical curves ($\delta$ small but still positive) and the critical one ($\delta=0$), the disorder slowly increases up to a parametrically large scale $L_r$ and then drops down. On the other side of the transition ($\delta<0$), the disorder shows the same behaviour for a while but eventually starts to increase rapidly, which reflects the crumpling. In  Fig.~\ref{Fig11} we plot the parameter $\delta$ as a function of the bare disorder.

%%%%%%%%%%%%%%%%%%%%%%%%%%%%%
%%%%%%%%%%%%%%%%%%%%%%%%%%%%%%

\subsection {Dynamic and static correlation functions }
\label{dyn-stat}

To characterize dynamic and static fluctuations in the membrane,
 we introduce  the following  correlation  functions: \cite{disorders-Morse-Grest}
\begin{eqnarray}
\overline{ \langle h^\alpha(0)   h^\beta(\mathbf x)  \rangle }&=&  \delta_{\alpha\beta} G^{d+s}(x)
 \label{gd+s-def}
  \\
  \overline{ \langle   h^\alpha(0) \rangle  \langle  h^\beta(\mathbf x) \rangle }&=& \delta_{\alpha\beta}G^{s}(x) ,
  \label{gs-def}
\end{eqnarray}
where angular brackets denote the Gibbs averaging, while the overline stands for the disorder averaging.
The function $G^{d+s},$ incorporates both dynamical and static fluctuations,  while $G^{s}$ includes static correlations only.  These functions depend on the absolute value of the distance $x=|\mathbf x|$,  so that their Fourier transforms
\begin{eqnarray}
G^{d+s}_{q}&=& \int  G^{d+s}(x)e^{-i\mathbf q\mathbf x} d^2\mathbf x \nonumber  \\& =&2\pi \int   G^{d+s}(x) J_0(qx) x dx ,
\\
G^{s}_{q}&=& \int  G^{s}(x)e^{-i\mathbf q\mathbf x}d^2\mathbf x\nonumber  \\& =& 2\pi \int    G^{s}(x) J_0(qx)x dx
\end{eqnarray}
depend on the absolute value of momentum $q=|\mathbf q|.$

The dynamic part of the fluctuations is thus given by the difference of these two functions,
 \be G^{d}_{ q}=
 G^{d+s}_{ q}-G^s_{ q}
 \ee
 (this function was used in the previous sections without index $d$). The correlation functions  defined above can be straightforwardly calculated on the basis of the above RG analysis.
Specifically, we first renormalize the theory from the original ultraviolet scale to the scale $1/q$.
As a result, all non-linear effects get incorporated in the renormalization of $\varkappa$ and $b$.
Having renormalized the couplings, we evaluate the correlation functions at the Gaussian level, which yields
 \BEA
 G^{d+s}_{q}&=&  \frac{1}{q^4}\left (  \frac{T}{{\varkappa}_q } +{b}_q \right), \label{gd+s}\\
 G^{d}_{ q}&=&   \frac{T}{{\varkappa}_q  q^4},  \label{gd}\\
 G^{s}_{\mathbf q}&=&  \frac{{b}_q}{q^4}, \label{gs}
   \EEA
   where ${\varkappa}_q$ and ${b}_q$ depend on $q$ according to  the RG equations derived above.

%%%%%%%%%%%%%%%%%%%%%%%%%
 \begin{figure}[ht]
\centerline{\includegraphics[width=0.45\textwidth]{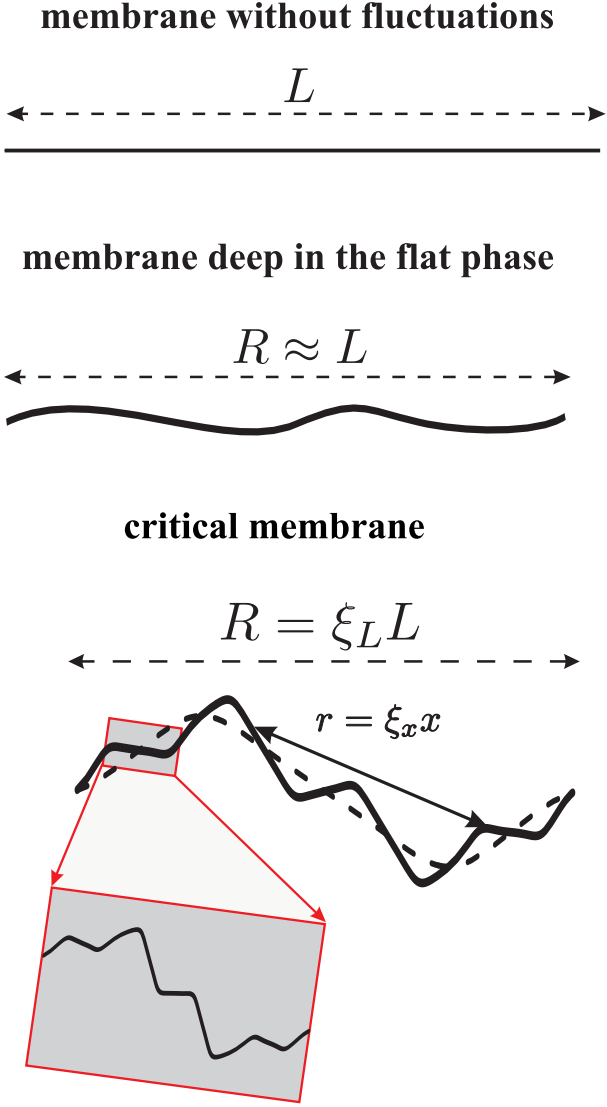} }
\caption{
{\it Top}: Membrane  with intrinsic size $L$ in the absence of  fluctuations; {\it Center}:    Membrane  deeply in the flat phase; {\it Bottom}: Membrane at criticality. The  linear size,  $R,$  of membrane with fluctuations  is smaller by  a factor $\xi_L.$  The distance, $r,$    in the embedding  space    between two points on membrane surface scales with the distance  on the reference plane with a local stretching factor: $r\simeq \xi_x x.$   Membrane shows self-similarity at different scales, so that the  dashed line represents  a coarse-grained shape of the membrane  at largest scale shown in the picture. The fractal structure of membrane is   illustrated  in the grey box magnifying a segment of the membrane.
}
\label{Fig12}
\end{figure}
%%%%%%%%%%%%%%%%%%%%%%%%%%%%%

At this point, it is worth recalling that the running scale $\Lambda$ for renormalization of   $\varkappa$ and $b$
is associated with a wave-vector $\mathbf q$ conjugated to the coordinate $\mathbf{x}$ in the reference plane.  From the experimental point of view, a more natural coordinate  on the membrane surface is given by   the  vector  $\mathbf r$ in the embedding space [see Eq.~\eqref{xi-new1}]. As seen from the   Fig.~\ref{Fig12},  a difference between  vectors $\mathbf r$  and $\mathbf x$   becomes stronger when the system approaches criticality.

  The physical correlation functions depend on the distance $r=|\mathbf r|$  in the embedding space between two points on the membrane surface:
 \BEA
 g^{d+s}(r)&=&  G^{d+s}\left(r/\xi\right)=\int   g^{d+s}_{ Q} J_0( Q r)\frac{Qd Q}{2\pi},
  \label{Gd+s-def}
 \\
  g^{s}(r)&=&  G^{s}\left(r/\xi\right)=\int   g^{s}_{ Q} J_0( Q r)\frac{Qd Q}{2\pi},
   \label{Gs-def}
 \\
  g^{d}(r)&=&g^{d+s}(r)-g^{s}(r).
  \EEA
 Here   $\xi=\xi_x$ is the running stretching parameter  that relates $r$ and $x:$  $r=\xi_x x$ [see Eq.~\eqref{xr}].
The correlation function  $g_{ Q}$ (measured experimentally) is related to $G_{ q}$
as follows:
\begin{equation}
g_{ Q/\xi } = \xi^2 ~G_{ Q} \ ,
\label{relationGg}
\end{equation}
with $\xi = \xi_{ Q }$.
In the flat phase, the rescaling factor $\xi$ remains finite in the infrared limit (i.e. the membrane linear size $R= \xi_L L$ in the embedding space is proportional to its intrinsic size $L$), and does not essentially affect the scaling.
Therefore, the difference between $g$ and $G$ is immaterial.
On the other hand, for a critical or near-critical membrane this difference is of crucial importance.  In particular, the RG flow is controlled by rescaled  couplings  $\tilde{\varkappa}$ and $\tilde{b}$ (rather than by $\varkappa$ and $b$.)  Further, we find from  Eq.~\eqref{relationGg} that the  correlation functions in the embedding space,   $g^{d+s}_{Q}$, $g^{d}_{ Q}$, and $g^{s}_{ Q},$ are given by Eqs.~\eqref{gd+s}, \eqref{gd},  and \eqref{gs} with $\varkappa$ and $b$ replaced by $\tilde{\varkappa}$ and $\tilde{b}$, respectively.

In the flat phase, we get the following asymptotic scaling behaviour of the static and dynamic correlation functions at $Q \to 0$:
 \BEA
  g^{d+s}_ Q \sim  g^{d}_ Q  & \propto&  \frac{1}{Q^{4-\eta}}   \;,
   \label{dynamic}
  \\
    g^{s}_ Q   &\propto&  \frac{1}{Q^{4-2\eta}} \;.
     \label{static}
 \EEA
 It is worth noting that an analogous asymptotic relation between the dynamic and static correlation functions was obtained in Ref.~\onlinecite{disorders-Morse-Grest} for $D=4-\epsilon$ for  the flat-phase  fixed point (called P4 point there).
 Equations \eqref{dynamic}, \eqref{static} imply, in particular, that the characteristic dynamic and static transverse
 excursions of a membrane  (root- mean-square values  of  the corresponding fluctuations of  $\mathbf h$) scale  with its size $R \propto L$ as follows:
 \BEA
 h_{{\rm rms}}^d &\propto & R^{1-\eta/2} \; , \\
   h_{{\rm rms}}^s &\propto & R^{1-\eta} \; .
  \EEA
At the crumpling transition point we have $\tilde \varkappa  = \varkappa_{\rm cr} = {\rm const}$,
 $\tilde b \propto L^{-\eta}$, and $\xi_L \propto L^{-\eta/2}$.
 Therefore, the characteristic magnitudes of the transverse excursions of the membrane scale at the transition as
 \BEA
  h_{{\rm rms}}^d &\propto & R \; , \label{hcr} \\
   h_{{\rm rms}}^s &\propto & R^{(2-2\eta)/(2-\eta)} \; ,
  \EEA
 and $R \propto L^{1-\eta/2}$.
  Since $\varkappa_{\rm cr} $ is proportional to $d_c^2$ [see Eq.~\eqref{cr1}]  the  dimensionless coefficient in  Eq.~\eqref{hcr} turns out to be  small, on the order of $1/\sqrt{d_c}.$

 \subsection {Ripple intensity and correlations}
It is natural to characterize
the intensity and  spatial correlations of ripples (i.e., of static transverse deformations)
with the static  dimensionless correlation function of spatial gradients of out-of-plane displacements. Such function can be expressed in terms of $g^s$  as follows
\BEA
\label{corr-spatial}
 H({r})&=&
 % \overline{  \nabla_{\mathbf{r}}  \langle \mathbf h (0) \rangle    \nabla_{\mathbf{r}} \langle \mathbf  h (\mathbf r/\xi) \rangle } \\
%\nonumber
% &=&
   \int \frac{d Q }{2\pi} Q^3 g^s_ Q   J_0( Q  r).
  \EEA
Using  Eq.~\eqref{relationGg}, we rewrite Eq.~\eqref{corr-spatial} as
\be
H({r})
=\int \frac{d^2\mathbf q}{(2\pi)^2} \frac{\tilde{b}_{q}}{q^2}  e^{i\mathbf q\mathbf x}.
\label{Hr}
\ee
It is worth noting that function $H(r)$  characterizes fluctuations and  correlations  of  normal vectors to the membrane surface. Such fluctuation of tilt angle of the surface are directly studied in graphene experiments, see below.

\begin{figure} [t]
\centerline{\includegraphics[width=0.45\textwidth]{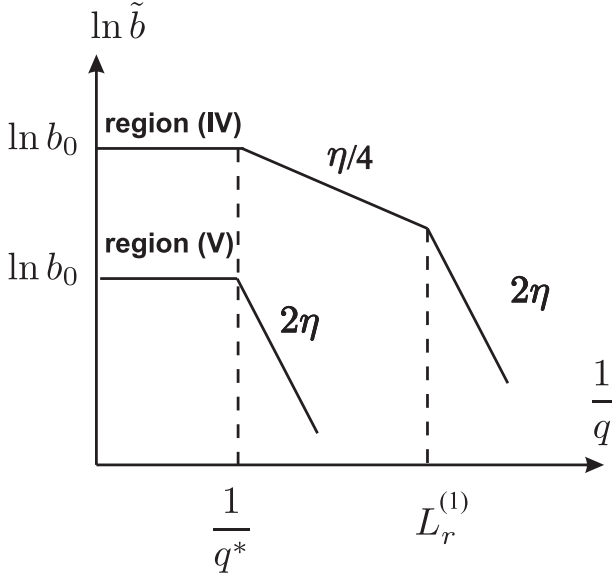} }
\caption{
Schematic  dependence of  the effective disorder strength $\tilde b_q$ on the length scale $q^{-1}$ on the log-log scale for regions (IV) and (V).
The indices $\eta/4$ and $2\eta$ denote the power-law decay exponents in the corresponding regimes.
}
\label{Fig13}
\end{figure}

Let us now discuss the temperature dependence of the size and height of the ripples deeply in the flat phase, i.e. in regions (IV) and (V) in Fig.~\ref{Fig7}. In this situation, the difference between $r$ and $x$ coordinates is not particularly important and can be discarded.
Using results of Secs.~ \ref{near}, \ref{away}, and \ref{spatial},
we  find  the behavior of $\tilde b_q$  (shown schematically in Fig.~\ref{Fig13})  and thus for $H(r)$ for these two cases.

%%%%%%%%%%%%%%%%%%%%%%%%
%%%%%%%%%%%%%%%%%%%%%

\subsubsection{ Strong    disorder [region (IV)].
}

The  calculation of the integral entering  Eq.~\eqref{Hr} yields
\be
H(r) \sim   \frac{b_0}{2\pi} \left \{\begin{array}{cc}
             %\displaystyle
             \ln  \left(\dfrac{1 }{q^* r} \right) , \quad & \!\!  r < 1/q^*,  \\[0.3cm]
             %\displaystyle
             \left(\dfrac{1}{q^*r}\right)^{\eta/4}  , \quad & \!\!\!\!\! 1/q^*<r < L_r^{(1)}, \\[0.3cm]
             %\displaystyle
             \left(\dfrac{1}{q^*L_r^{(1)}}\right)^{\eta/4} \left(\dfrac{L_r^{(1)}}{r}\right)^{2\eta} , \quad & \!\!   L_r^{(1)} <r,
          \end{array} \right.
\label{H3}
\ee
In Eq.~\eqref{H3},
\be
L_r^{(1)} \sim \frac{1}{q^*}~ e^{4 f_0/3\eta} =  \frac{1}{q^*}~ e^{4 b_0 \varkappa_0/3T \eta}
\label{Lr-near}
\ee
is the spatial scale  determined by the condition $f \sim 1 $.  For $r> L_r^{(1)},$  the exponent characterising the spatial decay of $H(r)$ changes because  of  stronger suppression of the disorder by increased bending rigidity.  Note that at very small distances, $r < a$ (where $a$ is the ultraviolet cutoff length, which is  of the order of the lattice constant), one should replace $r$ with $a$ in the first line of Eq.~\eqref{H3}.
We thus find that the dimensionless parameter $H(0)$ which represents the averaged squared surface tilt
is given by
\be
H(0) = \frac{b_0}{ 2\pi} \ln  \left (\frac{1}{q^* a} \right).
\label{HHH}
\ee
The length $1/q^*$ and the averaged squared tilt $H(0)$ are two natural parameters that characterise the characteristic extension and magnitude of ripples.
Importantly, both of them decrease with increasing temperature.

  \subsubsection{ Weak disorder  [region (V)].}
  \label{weak-disorder}

In the case of a weak disorder,     the only difference is that   disorder  falls with the exponent $2\eta$ from the very beginning of renormalization.
  Hence,  similar to the case of strong potential the characteristic ripple  size is determined  by  the  Ginzburg scale.  For $D=2$ this scale is determined from the condition $3{\cal N}  \Pi_\mathbf q \sim 1$ [see Eq.~\eqref{N}], yielding
\be
q^*=q_{{\cal N}}^*\simeq  \sqrt{\frac{3 A_2\mu   (2\mu +2\lambda) T}{(2\mu+\lambda) \varkappa^2}}.
\label{ginz}
\ee
Therefore, we find the ripple size
\be
L_r \simeq \frac{2\pi}{q^*}  \propto\frac{1}{\sqrt{T} }.
\label{Lrr}
\ee
The averaged squared surface tilt is  given by Eq.~\eqref{HHH} which can be rewritten as
\be
H(0)= \frac{b_0}{4\pi} \ln \left( \frac{T^*}{T}\right),
\label{H}
\ee
  where $T^* \simeq \varkappa^2(2\mu+\lambda)/ [6A_2\mu(\mu+\lambda) a^2] $.
  We see  that, also in this case,
    both the size of the ripples $L_r$ and the characteristic surface tilt $\sqrt{H(0)}$ decrease with increasing temperature.

\subsubsection{ Comparison with experiment}

Let us compare our results with available experimental data on suspended graphene. In Ref.~\onlinecite{Meyer} the parameters of ripples at room temperature were found to be $L_r \simeq 5 \div 10 \;{\rm nm}$ for the ripple size and  $\sqrt{H(0)}\simeq 5^\circ \simeq 0.1$ for the characteristic tilt angle.  Very similar results at $T=300\;{\rm K}$ were obtained in Ref.~\onlinecite{Kirilenko} which found $L_r \simeq 10 \;{\rm nm}$ and $H(0) \simeq 0.01$.  The small value of $H(0)$ indicates that the system is in the weak-disorder regime,  so that the results of Sec.~\ref{weak-disorder} are expected to be applicable. Indeed, the measured room-temperature ripple size $L_r$ agrees well with Eqs.~\eqref{Lrr} and \eqref{ginz} that yield $L_r\simeq 5 \;{\rm nm}$ at $T=300\;{\rm K}$.
Comparing \eqref{H} with the experimentally measured $H(0) \simeq 0.01$, we get an estimate for the disorder strength, $b_0 \simeq 0.03$.

The authors of Ref.~\onlinecite{Kirilenko} provided also some information about the temperature dependence of the ripple characteristics. Specifically, they performed measurements also at $T=150\;{\rm K}$ and found that, in comparison with the room temperature, the ripple size increased, $L_r \simeq 18 \;{\rm nm}$ whereas the averaged squared surface tilt $H(0)$ remained almost unchanged. These findings   are in reasonable agreement with our weak-disorder results \eqref{ginz} and \eqref{H} which predict a square-root increase of $L_r$ and a slow (logarithmic) increase of $H(0)$ with inverse temperature.

It is worth reminding the reader  at this point that static ripples coexist with dynamical fluctuations.
In the weak-disorder regime, the relative strength of the two types of fluctuations is controlled by the parameter $f_0 = b_0\kappa_0/T$, see Eqs.~\eqref{gd} and \eqref{gs}. According to the above estimate, this parameter is close to unity at room temperature for the samples experimentally studied in Refs.~\onlinecite{Meyer} and \onlinecite{Kirilenko}. However, it has a $1/T$ temperature dependence which reflects the fact that the dynamical fluctuations become suppressed with lowering temperature. Contrary to this, the typical tilt angle $\sqrt{H(0)}$ characterising static ripples gets enhanced with lowering temperature, Eq.~\eqref{H}. Therefore, measurement of the temperature dependence of $\sqrt{H(0)}$  may be useful for experimentally differentiating between the static and dynamic fluctuations.

 \section{Summary and outlook}
 \label{discussion}

In this article, we have discussed the rippling and the crumpling transition in  graphene with a static quenched disorder.      We derived RG equations, Eqs.~\eqref{RG-final-xi},  \eqref{RG-final-kappa}, and  \eqref{RG-final-f} for  a model of a crystalline membrane with out-of-plane (random curvature) disorder.
Equations \eqref{RG-final-kappa} and \eqref{RG-final-f} describe a combined flow of the running dimensionless bending rigidity $\tilde\varkappa/T$  and the running disorder strength $\tilde b \equiv f T/\tilde\varkappa$.  They yield, in particular, a critical curve $\tilde b(\tilde \varkappa)$ separating the flat and the crumpled phases, see Figs. \ref{Fig4}, \ref{Fig5}, and \ref{Fig6}. Equation  \eqref{RG-final-kappa}  controls the spatial contraction of the membrane due to its deformation.
Even deep in the flat phase,  random fluctuation of membrane tension caused by  the  disorder may  strongly affect  the behavior  of the bending rigidity $\tilde\varkappa_q/T$.  Specifically, for a sufficiently strong disorder, the bending rigidity decreases at the first stage of the renormalization,  reaches minimum,  and only  then starts to grow (Fig. \ref{Fig8}) . Furthermore, we have found that disorder $\tilde b_q$ also changes non-monotonously in the flat phase if the bare disorder $b_0$ is sufficiently strong.  Specifically, $\tilde b_q$ first increases slowly (logarithmically) with $L$,  then reaches a  maximum at  a certain scale $L_{ r} $, and finally  decreases according to a power law  at larger scales, see Figs. \ref{Fig9} and \ref{Fig10}.

The random static out-of-plane fluctuations of the graphene membranes  can be identified with experimentally observed ripples, with the length scale $L_{ r}$ playing the role of the characteristic ripple size. The found values and temperature dependencies of the ripples parameters---the size $L_r$ and the typical surface tilt angle $\sqrt{H(0)}$---are in a good agreement with experimental observations of Refs.~\onlinecite{Meyer,Kirilenko} if a disorder strength $b_0 \simeq 0.03$ is assumed.

 We have also  briefly discussed an in-plane disorder and showed that it is irrelevant in the RG sense (if one excludes a long-range disorder whose correlation function is highly singular at small momenta) and thus does not affect our main conclusions.  The effect of the in-plane disorder at atomic scales may, however, be important for determining the bare value of the out-of-plane disorder.

 Before closing the paper, we discuss some of possible directions of future research.

 \begin{enumerate}

 \item In our work we considered a free-standing membrane without tension. On the other hand, the tension may become essential under certain experimental conditions. In a clean case, such a membrane might demonstrate  a buckling transition.\cite{buck} It remains to be explored what will be the effect of disorder in this situation.

 \item We have assumed that the disorder is of short-range character. On the other hand, a finite density of topological defects (like dislocations or disclinations) may yield long-range-correlated disorder. An earlier work\cite{RLD1} predicts a variety of possible phases in a membrane with long-range disorder.  A study of crumpling transition and of rippling in a membrane with physically relevant long-range disorder remains an interesting prospect for future.

     \item Our analysis did not include terms preventing self-crossing of membrane which are known to become important in the crumpled phase. It remains to be investigated whether such terms may affect the physics in the near-critical regime.

\item  There is a certain analogy between the physics of a membrane and that of Anderson metal-insulator transition in disordered (and possibly interacting) systems\cite{evers08,finkelstein10}. In particular, the field theory of a disordered membrane developed above bears similarity with the $\sigma$-model description of the Anderson localization. Within this analogy, the flat phase corresponds to a metal, the crumpled phase to an insulator, and the  dimensionless bending rigidity $\tilde\varkappa/T$ to the dimensionless conductance. Remarkably, both problems manage to evade the Mermin-Wagner theorem, showing a transition also in $D=2$. (In the case of Anderson transition, this requires either spin-orbit coupling or electron-electron interaction.) Static fluctuations of local deformations (ripples) in a disordered membrane can be viewed a counterpart of mesoscopic fluctuations of wave functions (or local density  of states) in the Anderson-localization problem. An interesting and important question is whether this analogy can be pushed further and, in particular, whether the ripple statistics at the crumpling transition is characterised by multifractality that is a hallmark of the Anderson-transition critical point.

\item  On the experimental side, a more systematic study of rippling and crumpling in free-standing graphene  would be highly desirable. In particular, measurements of the temperature dependence of ripple parameters in a broader temperature range would
be of great interest. Furthermore, experiments on various kinds of membranes (including emerging 2D materials) are expected to be instrumental for exploring the whole phase diagram of the  problem.

 \end{enumerate}

 \section{Acknowledgements}

We thank L. Radzihovsky for useful discussions.
The research was funded by the Russian Science Foundation under the
grant No. 14-42-00044.

\appendix

\section{Applicability of the quasiclassical approximation}
\label{quasiclassical}

In this Appendix, we provide a justification for the quasiclassical approximation used in this paper and determine  the regime of its validity.

The quasiclassical  approximation  is valid for not too low temperatures  (see also Refs~\onlinecite{kats1,kats2,kats3} for discussion).
More specifically, it is well justified provided that temperature  is large compared to frequencies of both out-of-plane and in-plane phonons:  $T > \hbar \omega_\mathbf q,   $  $T > \hbar s q,$ where $\omega_\mathbf q$ is given by Eq.~\eqref{omega}, and $s=\sqrt{(2\mu+\lambda)/T}  $ is the velocity of the longitudinal in-plane phonons (Here we take account that velocity of transverse in-plane phonons  is smaller  for graphene  parameters.)      The  characteristic momentum $q$ of the discussed  problem is the Ginzburg scale $q_*$ which itself depends on temperature.
 According to Eq.~\eqref{ginz},  the condition $T> \hbar \omega_{q^*}$ can be rewritten as
 \be
 \rho \varkappa^3 > 36 A^2_2 \hbar^2 \frac{\mu^2 (\mu+\lambda)^2}{ (2\mu+\lambda)^2 }.
 \label{quasicl1}
 \ee
 Note that this inequality does not contain temperature.
 Substituting graphene parameters  ($\rho \simeq 7.6 \times 10^{-7}$ kg/m$^2,$
 $\lambda \simeq 3$ eV/\AA $^2,$  $\mu \simeq 9$ eV/\AA $^2,$  and $\varkappa \simeq 1$ eV),  we find that
l.h.s.   of this inequality  exceeds the r.h.s. by a  factor of the order of $10^3$, so that this requirement is perfectly met.    This result is not surprising, because the density  of graphene $\rho,$ entering this estimate is proportional to the atomic mass $M$ and therefore  is large   compared to typical electronic mass scales. In other words, the l.h.s. of Eq.~(\ref{quasicl1}) should be larger than its r.h.s. by a factor of the order of $M/m\sim 2\cdot 10^4$ (where $m$ is the electron mass).
  Hence,  for the problem discussed here, the  flexural phonons can be treated semiclassically for any temperature and with a very good precision. The corresponding criterium for longitudinal phonons reads
\be
T > T_{\rm in} = 6 A_2  \frac{\hbar^2 \mu (\mu+\lambda)}{\rho \varkappa^2}.
 \label{quasicl2}
\ee
The r.h.s. of this inequality can be estimated as $(m/M)E_a$, where $E_a$ is a characteristic atomic energy scale. Taking $E_a=10\;$eV, we get a rough estimate $T_{\rm in} \simeq 5\;$K. Using known results for the mechanical parameters of graphene yields a somewhat larger value,
$T_{\rm in} \simeq 80$ K, which still leaves enough room for the validity of the semiclassical theory.   Furthermore, our theory remains applicable also at lower temperatures,  $T< T_{\rm in}$, where it describes the physics on sufficiently large spatial scales,   $q \lesssim  T/\hbar s$.

\section{Screening of $h^4$ interaction}
\label{technical}

In this Appendix, we present technical details of calculation of the screening.
As a starting point we use  equation  for free energy  derived in  Ref.~\onlinecite{Doussal}
\BEA
\frac{F}{T}&=& \frac{\varkappa}{2T}\int (dk) k^4 |\mathbf h_\mathbf k|^2 + \frac{1}{4d_c}\int (dk_1 dk_2 dk_3) R_{\alpha\beta\gamma\delta} (\mathbf q)  \nonumber \\
&\times& k_{1\alpha}k_{2\beta}k_{3\gamma}k_{4\delta}  (\mathbf h_{\mathbf k_1}  \mathbf h_{\mathbf k_2})  (\mathbf h_{\mathbf k_3}  \mathbf h_{\mathbf k_4}),
\label{Frad}
\EEA
where $\mathbf q=\mathbf k_1+ \mathbf k_2$  and $\mathbf k_1+ \mathbf k_2+\mathbf k_3+ \mathbf k_4=0,$
and the interaction kernel reads
\BEA \label{Rabcd}
R_{\alpha \beta \gamma \delta} (\mathbf q) &=&   \frac{{\cal N}}{D-1}  P_{\alpha \beta}P_{\gamma \delta}
\\
\nonumber
&+&{\cal M} \left(\frac{P_{\alpha \gamma}P_{\beta \delta}+P_{\alpha\delta}P_{ \beta \gamma}}{2} -\frac{P_{\alpha \beta}P_{\gamma \delta}}{D-1} \right). \EEA
Choosing $\mathbf k_1=\mathbf k+\mathbf q,$ $\mathbf k_2=-\mathbf k,$ $\mathbf k_3=-\mathbf k'-\mathbf q,$ $\mathbf k_4=\mathbf k'$, we find, after some algebra,
\be R_{\alpha\beta\gamma\delta} (\mathbf q)
 k_{1\alpha}k_{2\beta}k_{3\gamma}k_{4\delta} = R_\mathbf q (\mathbf k, \mathbf k'),
 \ee
where $R_\mathbf q (\mathbf k, \mathbf k')$ is given by Eq.~\eqref{K} of the main text.    Screened   interaction $\tilde{R}_{\alpha\beta\gamma\delta}$ obeys\cite{Doussal}
\be \label{Rabcd-screened}
\tilde{R}_{\alpha\beta\gamma\delta}={R}_{\alpha\beta\gamma\delta}- {R}_{\alpha\beta\gamma'\delta'}\Pi_{\gamma'\delta'\alpha'\beta'}\tilde{R}_{\alpha'\beta'\gamma\delta},
\ee
with the tensor polarization operator  $\Pi_{\gamma\delta\alpha\beta}$ given by
\be
\Pi_{\gamma\delta\alpha\beta}=\int (dk)~ k_\alpha k_\beta k_\gamma k_\delta G^0_\mathbf kG^0_{\mathbf q-\mathbf k}.
\label{Piabcd}
\ee
Because of the rotation symmetry one can write
\BEA
&&\Pi_{\gamma\delta\alpha\beta}= (\delta_{\alpha\beta}\delta_{\gamma\delta} +\delta_{\alpha\gamma}\delta_{\beta\delta}+\delta_{\alpha\delta}\delta_{\beta \gamma})~\Pi_\mathbf  q
\nonumber
\\
&&+  (\delta_{\alpha\beta}~ q_{\gamma} q_{\delta} + \cdots  ) ~\Pi^{(1)}_\mathbf  q +q_{\alpha} q_{\beta} q_{\gamma} q_{\delta} ~\Pi^{(2)}_\mathbf  q,
\EEA
where  functions  $\Pi_\mathbf  q,$ $\Pi^{(1)}_\mathbf  q,$  and $\Pi^{(2)}_\mathbf  q$ depend on $|\mathbf q|$ only, and
$(+\cdots)$  stands for sum over  permutations of $\alpha,~ \beta,~\gamma,$   and $\delta.$  Due to the  projection operators $\hat P$ entering in the unscreened coupling, Eq.~\eqref{Rabcd},  functions  $\Pi^{(1)}_\mathbf  q$  and $\Pi^{(2)}_\mathbf  q$  drop out from  Eq.~\eqref{Rabcd-screened}.    The function   $\Pi_\mathbf  q$ can be easily obtained by multiplication of Eq.~ \eqref{Piabcd} by $P_{\alpha\beta}P_{\gamma\delta}$ and summation over repeated indices. Taking into account that trace of the matrix $P_{\alpha\beta}$ equals to $D-1,$ we arrive at Eq.~\eqref{Pi} of the main text. Further, substituting Eq.~\eqref{Piabcd} into Eq.~\eqref{Rabcd-screened}, we find that the screened interaction $\tilde{R}_{\alpha\beta\gamma\delta}$  can be written  in the same form as  Eq.~\eqref{Rabcd} but with ${\cal N}$ and ${\cal M}$ replaced by  their screened values  ${\cal N}_\mathbf q$ and ${\cal M}_\mathbf q$. These values are given  by  Eqs.~\eqref{N}  and \eqref{M} of the main text, respectively.

\vskip0cm

\section{Regularization integrals}\label{regul}

In order to shed light on a connection of our approach to the  SCSA, we employ the regularization  integrals $A(D,\eta)$ and $B(D,\eta)$ introduced in Ref.~\onlinecite{Doussal}:
\begin{eqnarray} \label{A}
  A(D,\eta)&=& \frac{1}{D^2-1} \int (dx) \frac{x_\perp^4}{x^{4-\eta}|\mathbf x - \mathbf n|^{4-\eta}} \\
   &=&\frac{\Gamma\left( \frac{D+\eta}{2}\right) \Gamma\left(\frac{ 4-2\eta-D}{2}\right)}{2^{2D+\eta+1} ~\pi^{(D-1)/2}~\Gamma^2\left( \frac{4-\eta}{2}\right)\Gamma\left( \frac{1+D+\eta}{2}\right)},\nonumber
\end{eqnarray}
\begin{eqnarray} \label{B}
  B(D,\eta)&=&  \int (dx) \frac{x_\perp^4}{x^{D+2\eta}|\mathbf x - \mathbf n|^{4-\eta}} \\
   &=&\frac{(D^2-1)\Gamma\left( \frac{D+\eta}{2}\right) \Gamma\left(\frac{ \eta}{2}\right) \Gamma(2-\eta)}{2^{2+D} ~\pi^{D/2}~\Gamma\left( \frac{4-\eta}{2}\right)\Gamma\left( \frac{D+2\eta}{2}\right)\Gamma\left( \frac{4+D-\eta}{2}\right)}.
\nonumber
\end{eqnarray}
Here $\mathbf n$ is an arbitrary unit-length vector and $\mathbf x_\perp=\mathbf x- \mathbf n(\mathbf x \mathbf n).$
These integrals naturally arise  when one uses for calculation of the polarization loop the Green function, Eq.~\eqref{G}, with the self-energy found self-consistently by replacement of the bare Green function $G_\mathbf k^0$ in Eq.~\eqref{Sigma}   with     $G_\mathbf k.$
Evaluation of both integrals can be performed in the following way.    First, one uses the identity
\BEA
&&\frac{1}{x^\theta |\mathbf x-\mathbf n|^{4-\eta}}= \frac{1}{\Gamma\left ( \frac{\theta}{2}\right)\Gamma\left ( \frac{4-\eta}{2}\right)}
\nonumber
\\
&&\times\int\limits_0^\infty dt_1\int\limits_0^\infty dt_2 ~t_1^{\theta/2-1} t_2^{1-\eta/2} e^{- t_1 x^2 -t_2 (\mathbf x-\mathbf n)^2},
\nonumber
\EEA
 where $\theta=4-\eta$  for the integral  $A$   and $ \theta=D+2\eta$  for the integral $B$.  The integral over $dx=d\mathbf x^D/(2\pi)^D$ becomes then a Gaussian one and is easily calculated.  The remaining integral  can be done  by  using the following change of variables:      $t_1= \tau  e^z \cosh z  $ and $t_2= \tau e^{-z}\cosh z$.     After lengthy but straightforward calculations one arrives at  Eqs.~\eqref{A} and \eqref{B}.


\begin{thebibliography}{99}
\bibitem{foot-Gazit} The stability of the SCSA with respect to higher-order contributions
was addressed in Ref.~\onlinecite{Gazit1}, where the inclusion of second-order corrections
in the parameter $1/d_c=1$ was found to yield $\eta\approx 0.789$ for physical
membranes. The SCSA has been applied to graphene in Ref.~\onlinecite{kat2}.
\bibitem{foot-xi} As follows from Eqs.~\eqref{dxi} and \eqref{hhh},  renormalization of  $\xi$ starts   from  the inverse ultraviolet cutoff length, $1/a.$    Here we neglect renormalization of $\xi$  in the interval   $q^* < q <1/a $  assuming that  $(1/d_c)\ln(1/q^*a) \ll1.$
\bibitem{foot-Ginz} The Ginzburg length as a characteristic scale for thermally excited ripples was also
found in Ref.~\onlinecite{Hasselmann}. Numerical simulations of graphene flakes performed in Ref.~\onlinecite{Fasolino}
also yielded the spatial scale of ripples close to the Ginzburg length.
In Refs. \onlinecite{San-Jose} and \onlinecite{Gazit2}, the size
of ripples was related to the coupling of phonons with electrons, yielding a $T$-independent length
(ultraviolet atomic scale). Interestingly, taking into account the numerical factors, this scale was estimated~\cite{Gazit2}
to be close to the thermal Ginzburg scale corresponding to the room temperature. We emphasize that the membrane
fluctuations discussed in the above works are dynamical rather than static, see Sec. \ref{dyn-stat}.
\end{thebibliography}
\end{document}